\documentclass[]{interact}

\usepackage[numbers,square,sort&compress,super]{natbib}
\usepackage{url}
\usepackage{orcidlink}

\usepackage{array}
\newcolumntype{R}[1]{>{\raggedright\let\newline\\\arraybackslash\hspace{0pt}}p{#1}}

\usepackage{enumitem}

\usepackage{multirow}
\usepackage{multicol}

\usepackage{lmodern}

\begin{document}

\title{On the reproducibility of discrete-event simulation studies in health research: an empirical study using open models}

\author{
\name{Amy Heather\textsuperscript{a}\orcidlink{0000-0002-6596-3479}, Thomas Monks\textsuperscript{a}\orcidlink{0000-0003-2631-4481}\thanks{CONTACT Thomas Monks. Email: t.m.w.monks@exeter.ac.uk}, Alison Harper\textsuperscript{b}\orcidlink{0000-0001-5274-5037}, Navonil Mustafee\textsuperscript{b}\orcidlink{0000-0002-2204-8924}, Andrew Mayne\textsuperscript{c}\orcidlink{0000-0003-1263-2286}}
\affil{\textsuperscript{a}University of Exeter Medical School, St Luke's Campus, Heavitree Rd, Exeter, UK;
\textsuperscript{b}Centre for Simulation, Analytics and Modelling (CSAM), University of Exeter Business School, Streatham Campus, Rennes Drive, Exeter, UK;
\textsuperscript{c}Somerset NHS Foundation Trust, Taunton, UK}
}

\maketitle

\textbf{Abstract word count:} 178

\textbf{Article word count:} 8013

\newpage

\begin{abstract}
Reproducibility of computational research is critical for ensuring transparency, reliability and reusability. Challenges with computational reproducibility have been documented in several fields, but healthcare discrete-event simulation (DES) models have not been thoroughly examined in this context. This study assessed the computational reproducibility of eight published healthcare DES models (Python or R), selected to represent diverse contexts, complexities, and years of publication. Repositories and articles were also assessed against guidelines and reporting standards, offering insights into their relationship with reproducibility success. Reproducing results required up to 28 hours of troubleshooting per model, with 50\% fully reproduced and 50\% partially reproduced (12.5\% to 94.1\% of reported outcomes). Key barriers included the absence of open licences, discrepancies between reported and coded parameters, and missing code to produce model outputs, run scenarios, and generate tables and figures. Addressing these issues would often require relatively little effort from authors: adding an open licence and sharing all materials used to produce the article. Actionable recommendations are proposed to enhance reproducibility practices for simulation modellers and reviewers.
\end{abstract}


\section{Introduction}

The reproducibility and replicability of published research are widely recognised as cornerstones of rigorous science and have been investigated across numerous disciplines.\cite{baker_reproducibility_2020} Within the field of modelling and simulation, the importance of reproducibility has gained increasing attention and is recognised as essential for ensuring the reliability and impact of simulation-based research.\cite{monks_strengthening_2019, fisar_reproducibility_2024, wilsdorf_automatic_2023} Despite this growing emphasis, no study has empirically assessed the reproducibility of results generated by healthcare discrete-event simulation (DES) models. These models are important tools in healthcare decision-making, aiming to enhance patient outcomes and optimise health service delivery by addressing challenges such as patient flow management, resource allocation, and cost-effectiveness analyses.\cite{salleh_simulation_2017} This study investigates the reproducibility of open DES models - those with code made publicly available for others to run and reuse - with an emphasis on identifying barriers and facilitators to reproduction. The study identifies the work required to reuse the author-supplied model code to reproduce the simulation results presented in the tables, charts, and text of journal articles, including any troubleshooting required during the process. The study aims to provide actionable guidance to simulation modellers to improve the reproducibility of their work.

\subsection{Why reproducibility matters}

Reproducibility is the ability to regenerate published results using the provided code and data. Failures in reproducibility may indicate underlying issues, such as incorrect parameters, missing or outdated code, or unexpected changes in software behaviour due to updates. Achieving reproducibility fosters trust and transparency, building confidence that the model behaves consistently with the original implementation. It is vital to establish reproducibility before a model is reused.\cite{sandve_ten_2013, alston_beginners_2021, harper_facets_2021}

\begin{figure}
    \centering
    \resizebox*{14cm}{!}{\includegraphics{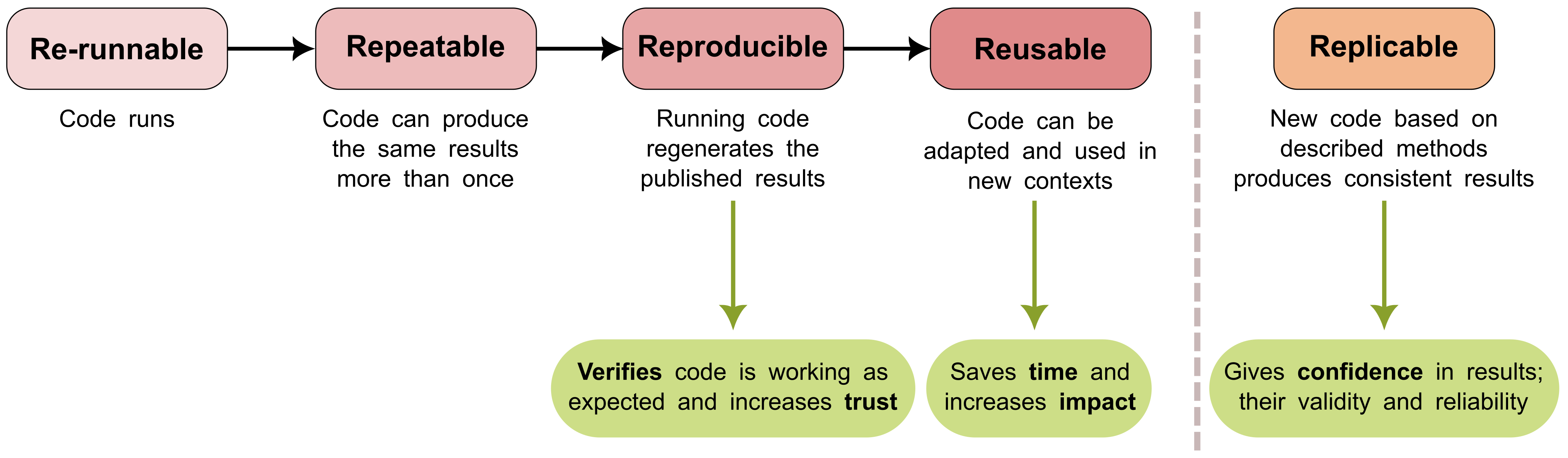}}
    \caption{Five standards for scientific code, as described in Benureau et al. (2018).\cite{benureau_re-run_2018}}
    \label{figure-5r}
\end{figure}

Reuse is the adaptation and application of a model in new contexts, amplifying the potential real-world impact of the model. Reuse is in high demand; for example, 37\% of Wellcome Trust-funded researchers report using external code, often from colleagues or community repositories.\cite{eynden_survey_2016} As shown in Figure \ref{figure-5r}, reproducibility exists within a continuum of code attributes, including being re-runnable, repeatable, reusable and replicable. Together, these characteristics underpin the validity and reliability of simulation models.\cite{benureau_re-run_2018}

Reproducibility also benefits authors by making it easier to revisit and reuse their own code, such as for updating outputs or conducting new analyses.\cite{alston_beginners_2021} Troubleshooting non-reproducible code can be time-consuming or even impossible if required information is now lost.\cite{sandve_ten_2013, alston_beginners_2021} For instance, code may need to be re-run following peer review.\cite{alston_beginners_2021} Given that the mean time from submission to publication in biomedical research ranges from 27 to 639 days,\cite{andersen_time_2021} it is important that code remains functional and reproducible long after its initial execution. Even if authors retain access to their code, failure to record exact parameters for every scenario or document the computational environment used, for example, could make the code non-reproducible when revisited, even if only a few months later. Finally, reproducibility often improves overall code quality. It encourages clear structure and documentation, reducing errors and ambiguities.\cite{alston_beginners_2021} These advantages apply regardless of whether code is shared publicly or remains proprietary.

\subsection{Addressing challenges in reproducibility}

The computational reproducibility of published research has been investigated across numerous disciplines. These studies, which focus on the reproducibility of results using shared code and data, vary in scope and methodology - for example, the extent of troubleshooting allowed when running code. Across many of these studies though, a consistent finding is that the majority of attempts to reproduce results fail, with examples in epidemiology,\cite{henderson_reproducibility_2024} public health,\cite{wood_push_2018} economics,\cite{galiani_incentives_2017, mccullough_lessons_2006} political science,\cite{eubank_lessons_2016} physics,\cite{krafczyk_learning_2021, stodden_enabling_2018} and hydrology.\cite{stagge_assessing_2019}

However, some studies do observe higher rates of reproducibility, often under specific conditions. Investigations in political science,\cite{stockemer_data_2018} psychology,\cite{obels_analysis_2020} and articles published in \textit{Science}\cite{stodden_empirical_2018} found that while many papers lack accessible data or code, the majority of results could be reproduced in those that provide these resources, in these cases. Similarly, studies in geoscience\cite{konkol_computational_2019} and a large-scale analysis of articles with Jupyter notebooks\cite{samuel_computational_2024} found that artefacts may often fail to execute, but when they do, the majority of results could be reproduced. High reproducibility was observed in management science,\cite{fisar_reproducibility_2024} where 95\% of studies were fully or largely reproduced - though this dropped to 68\% when including studies with inaccessible datasets.

Journals can play a pivotal role in addressing reproducibility issues, as summarised in Table \ref{table-iniatives}. Simple initiatives include requiring code availability statements or requiring that code is shared upon request. However, authors may simply state that code is unavailable or choose not to share it.\cite{stodden_empirical_2018, janssen_code_2020} For example, in a study of individual- and agent-based simulation models, researchers emailed authors who had stated in journal articles that their code was available upon request. However, they received responses from less than 1\% of authors - some provided links to the code, but others stated they no longer had access to it.\cite{janssen_code_2020} To address this, journals may mandate code sharing via public repositories or archives.\cite{association_for_computing_machinery_acm_artifact_2020, blohowiak_badges_2023, institute_of_electrical_and_electronics_engineers_ieee_about_2024, cadwallader_collaborating_2021, loder_mandatory_2024, fisar_reproducibility_2024, nature_computational_science_code_2023, hardwicke_transparency_2024} However, code sharing alone does not guarantee reproducibility, so journals may also review repositories to check that code is documented, complete and/or likely to be executable.\cite{management_science_code_2019, hardwicke_transparency_2024, nature_computational_science_code_2023, association_for_computing_machinery_acm_artifact_2020, institute_of_electrical_and_electronics_engineers_ieee_about_2024} Some journals attempt to run the provided code during peer review to assess its computational reproducibility,\cite{hardwicke_transparency_2024, nature_human_behaviour_promoting_2024, association_for_computing_machinery_acm_artifact_2020, institute_of_electrical_and_electronics_engineers_ieee_about_2024} or accept reproductions of published work by other groups as formal publications.\cite{nature_machine_intelligence_revisiting_2022} When code or data cannot be shared for ethical or legal reasons, at least providing the reviewer access for reproducibility checks is valuable, as others will not be able to interrogate it and assess validity after publication.\cite{eubank_lessons_2016} Alternatively, authors could share code with synthetic data for validation.\cite{monks_strengthening_2019} To provide a clear overview and comparison of journal policies, the Centre for Open Science (COS) reviews them on their website (\url{https://topfactor.org/}) against their Transparency and Openness Promotion (TOP) Guidelines.\cite{centre_for_open_science_top_2024}

Several conferences have also taken similar actions. For example, the \textit{Association for Computing Machinery (ACM) Special Interest Group on Simulation and Modelling (SIGSIM) Principles of Advanced Discrete Simulation (PADS)} conference has a reproducibility committee to assess papers against the ACM badges.\cite{acm_sigsim_reproducibility_2025}

\begin{table}
\tbl{Cross-disciplinary journal initiatives to improve computational transparency and reproducibility.}
{\begin{tabular}{R{3cm} R{10cm}} \toprule
\textbf{Initiatives} & \textbf{Examples} \\ \midrule

Mandatory code sharing & 
\begin{minipage}[t]{\linewidth}
Journals incentivise or require code to be shared publicly through:
\begin{itemize}[nosep, wide=0pt, leftmargin=*, after=\strut]
    \item Policies for mandatory code sharing.\cite{cadwallader_collaborating_2021,loder_mandatory_2024,fisar_reproducibility_2024, management_science_code_2019,nature_computational_science_code_2023,hardwicke_transparency_2024}
    \item Badges awarded to articles with shared code.\cite{association_for_computing_machinery_acm_artifact_2020,blohowiak_badges_2023,institute_of_electrical_and_electronics_engineers_ieee_about_2024}
\end{itemize} \end{minipage} \\ \addlinespace

Review of shared artefacts & 
\begin{minipage}[t]{\linewidth}
Journals evaluate the shared artefacts with:
\begin{itemize}[nosep, wide=0pt, leftmargin=*, after=\strut]
    \item Policies requiring details on the language and packages, commented code, and test data/problems sufficient for replication.\cite{management_science_code_2019}
    \item Policies requiring open licenses and clear documentation.\cite{hardwicke_transparency_2024}
    \item Integration with sharing platforms which streamline submission and peer review of code within the publication workflow.\cite{nature_computational_science_code_2023}
    \item Badges awarded following evaluation of artefacts (e.g. checking they are complete, executable and documented).\cite{association_for_computing_machinery_acm_artifact_2020,institute_of_electrical_and_electronics_engineers_ieee_about_2024}
\end{itemize} \end{minipage} \\ \addlinespace

Assess computational reproducibility &
\begin{minipage}[t]{\linewidth}
Journals actively test whether results can be reproduced through:
\begin{itemize}[nosep, wide=0pt, leftmargin=*, after=\strut] 
    \item Collaboration with independent institutions to reproduce and/or replicate a sample of published articles from the journal.\cite{hardwicke_transparency_2024, nature_human_behaviour_promoting_2024}
    \item Independent reproduction attempts during peer review, with badges awarded for successful reproductions.\cite{association_for_computing_machinery_acm_artifact_2020,institute_of_electrical_and_electronics_engineers_ieee_about_2024,hardwicke_transparency_2024}
\end{itemize} \end{minipage} \\ \addlinespace

Accept reproductions as formal publications &
\begin{minipage}[t]{\linewidth}
Journals recognise reproduction work as scholarly contribution by publishing:
\begin{itemize}[nosep, wide=0pt, leftmargin=*, after=\strut] 
    \item Independent reproduction studies from independent groups reproducing, reusing, or extending published code.\cite{nature_machine_intelligence_revisiting_2022}
    \item Reports documenting reproduction attempts during peer review as supplementary material.\cite{association_for_computing_machinery_acm_artifact_2020}
\end{itemize} \end{minipage} \\ \bottomrule

\end{tabular}}
\label{table-iniatives}
\end{table}

\subsection{Healthcare simulation models}

This study focuses on healthcare simulation models, specifically DES. DES has a wide range of applications, including healthcare operations and system design, medical decision-making, and infectious disease modelling.\cite{salleh_simulation_2017} It is the most commonly used simulation method in healthcare.\cite{philip_simulation_2023, roy_healthcare_2021, salleh_simulation_2017, salmon_structured_2018}

In recent years, efforts have been made to enhance the sharing and reuse of healthcare simulation models. The \textit{Strengthening The Reporting of Empirical Simulation Studies-DES} (STRESS-DES) guidelines from Monks et al. (2019)\cite{monks_strengthening_2019} were developed to facilitate replication by providing technical details necessary to recreate a model. Zhang et al. (2020)\cite{zhang_reporting_2020} developed a ``generic reporting checklist" focused on model quality, derived from the \textit{International Society for Pharmacoeconomics and Outcomes Research (ISPOR)–Society for Medical Decision Making (SMDM) Modelling Good Research Practices Task Force} reports. In 2024, the \textit{Sharing Tools and Artefacts for Reusable Simulations} (STARS) framework was introduced to provide guidance on which artefacts should be shared in repositories to facilitate model reuse.\cite{monks_towards_2024}

These guidelines and frameworks (fully detailed in Appendix \ref{appendix-criteria}) were developed because many healthcare DES models are not shared in ways that facilitate understanding, use, and reproducibility. A 2023 review\cite{monks_computer_2023} of 564 healthcare DES papers (2019-2022) revealed only 8.3\% shared model code, although this was improving over time. Among shared models, only 60\% included a README, with just 32\% explaining how to run the model. Less than half provided an open licence, and only 10.6\% archived their models. Across all papers, only 12.8\% used reporting guidelines.\cite{monks_computer_2023}

There are no studies specifically focused on the reproducibility of healthcare DES models. However, several studies have included simulation models in their sample, though these were in other domains.\cite{fisar_reproducibility_2024, eubank_lessons_2016, krafczyk_learning_2021, stodden_empirical_2018} One healthcare-specific example is Henderson et al. (2024)\cite{henderson_reproducibility_2024} which examined the reproducibility of infectious disease models, and included some simulation models. They analysed two samples of 100 articles and found that 21 to 23\% were fully reproducible, 42 to 46\% were partially reproducible, and 31 to 37\% were not reproducible, with minimal troubleshooting allowed.\cite{henderson_reproducibility_2024} However, their study did not explore the barriers and facilitators of reproducibility within this context, an aspect that would have required more extensive troubleshooting. While the non-healthcare simulation studies have explored barriers and facilitators,\cite{fisar_reproducibility_2024, eubank_lessons_2016, krafczyk_learning_2021, stodden_empirical_2018} they were not limited to simulation models or healthcare contexts.

This study aims to assess the reproducibility of published healthcare DES models by examining a sample of eight studies. It will identify factors that facilitate or hinder each reproduction and will evaluate adherence to relevant frameworks, badges and reporting guidelines, to understand how these influence reproducibility. By understanding which elements are most important when reporting results in articles and sharing code, the study seeks to identify key factors that can improve reproducibility in healthcare DES models. Rather than providing a theoretical list of best practices, this empirical approach focuses on understanding what modellers actually do in practice and what specific barriers arise when attempting reproduction, enabling more targeted and actionable guidance than general principles alone. While not all studies may have been designed with computational reproducibility in mind, it is increasingly recognised as an important consideration for any research involving code,\cite{benureau_re-run_2018} as it helps verify the integrity of findings. Adopting basic practices to improve reproducibility can enhance scientific rigour and is achievable without requiring advanced technical skills.

\section{Methods}

\begin{figure}[!htp]
    \centering
    \resizebox*{14cm}{!}{\includegraphics{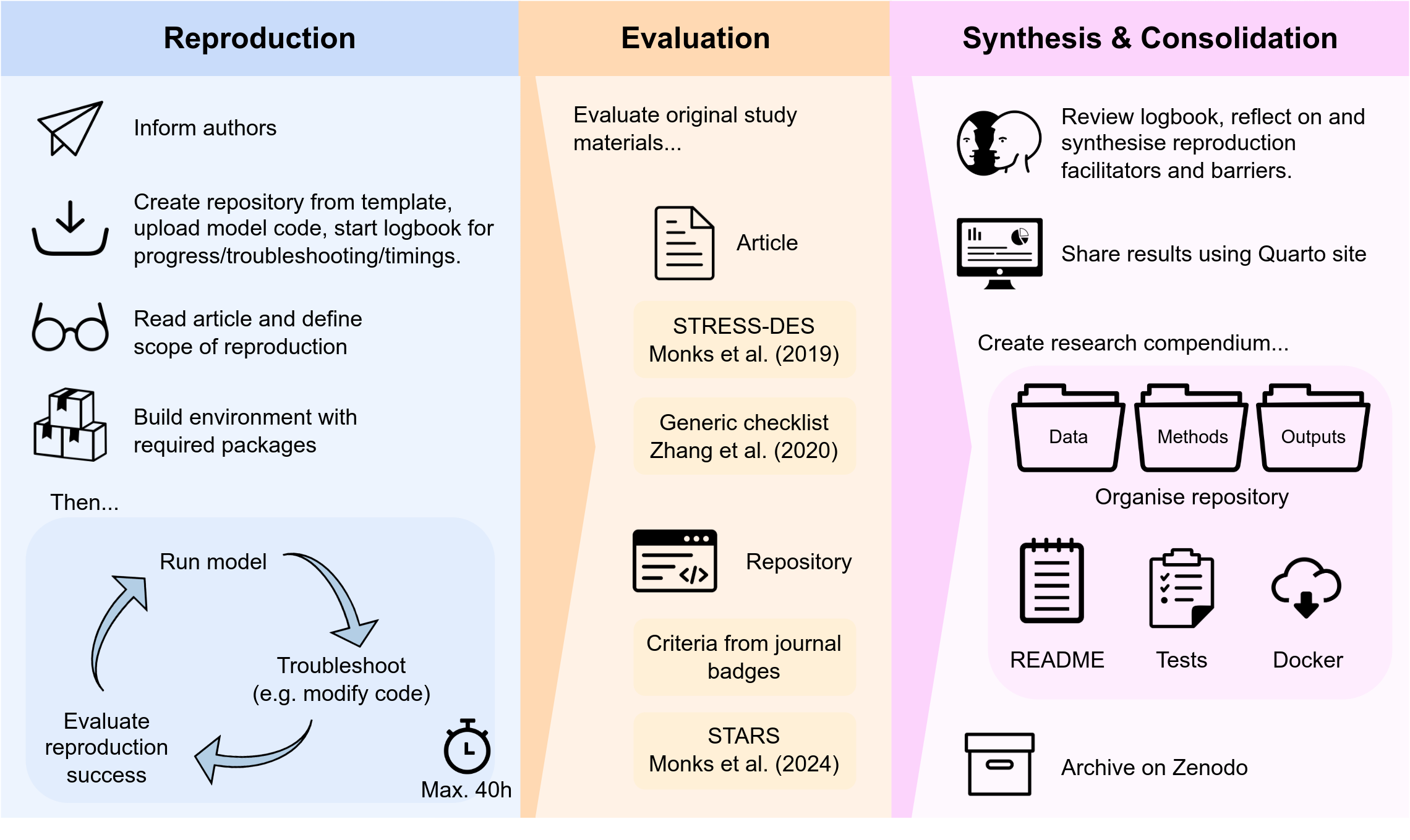}}
    \caption{Study methodology.\\ Abbreviations: STARS, \textit{Sharing Tools and Artefacts for Reusable Simulations}; STRESS-DES, \textit{Strengthening The Reporting of Empirical Simulation Studies-Discrete-Event Simulation}.}
    \label{figure-method}
\end{figure}

The study protocol (as summarised in Figure \ref{figure-method} and described below) was pre-registered on 20 June 2024.\cite{heather_protocol_2024} It was informed by prior reproduction studies,\cite{krafczyk_learning_2021, wood_push_2018, schwander_replication_2021, laurinavichyute_share_2022, konkol_computational_2019} and refined through a pilot on a Python DES model by Allen et al. (2020).\cite{allen_simulation_2020}

\subsection{Sample}

This study included eight models which were selected through purposive sampling. Of these, seven models were drawn from the sample of an existing systematic scoping review (2018-2022),\cite{monks_computer_2023} whilst one additional model was identified through an informal exploratory search of earlier literature. Models were developed in Python \cite{python_core_team_python_2024} or R\cite{r_core_team_r_2024}, the most popular Free and Open Source Software (FOSS) languages for healthcare DES.\cite{monks_computer_2023} Models were selected to ensure diversity across a range of factors including the health focus (e.g. healthcare condition, specific system), geographical context, and model complexity - as in Table \ref{table-studies}.

The protocol initially planned to include six studies, but this was later expanded to eight during the research process. This allowed the inclusion of another ``medium-sized" model \cite{wood_value_2021} that was felt to better represent typical operational research applications of DES in health compared to some larger-scale models. An older model\cite{hernandez_optimal_2015} was also included, enabling consideration of reproducibility challenges from an older code base (approximately a decade ago), as opposed to more recent studies (with all others published within the last five years).

\begin{table}[!ht]
\tbl{Description of the included studies.}
{\begin{tabular}{%
    R{1.7cm}%
    R{2cm}%
    R{4.2cm}%
    R{1.5cm}%
    R{1.1cm}%
    R{2.5cm}} \toprule
\textbf{Study} & \textbf{Journal/ Conference} & \textbf{Model} & \textbf{Geographical context} & \textbf{Scope*} & \textbf{Language**} \\ \midrule

Hernandez et al. (2015)\cite{hernandez_optimal_2015} &
\textit{Computers \& Industrial Engineering} &
Optimal staff numbers at sites dispensing counter-measures in a public health emergency &
New York, US &
8 &
Python (\textit{SimPy})\cite{team_simpy_simpy_2024} \& R \\ \addlinespace

Huang et al. (2019)\cite{huang_optimizing_2019} &
\textit{Frontiers in Neurology} &
Wait time and resource utilisation of an endovascular clot retrieval service under different configurations &
Australia &
8 &
R (\textit{simmer})\cite{ucar_simmer_2019} \\ \addlinespace

Lim et al. (2020)\cite{lim_staff_2020} &
\textit{Clinical Biochemistry} &
Impact of staff configurations and safety measures on COVID-19 transmission in a laboratory &
Not specified &
9 &
Python \\ \addlinespace

Kim et al. (2021)\cite{kim_modelling_2021} &
\textit{PLOS ONE} &
Impact of COVID-19-related disruption on outcomes of an abdominal aortic aneurysm screening programme &
England &
10 &
R \\ \addlinespace

Johnson et al. (2021)\cite{johnson_cost_2021} &
\textit{Applied Health Economics and Health Policy} &
Cost-effectiveness of different primary care-based case detection strategies for COPD &
Canada &
5 &
R (interface for C++ model)   \\ \addlinespace

Wood et al. (2021)\cite{wood_value_2021} &
\textit{Medical Decision Making} &
Deaths and life years lost under different triage strategies for an intensive care unit during COVID-19 &
UK &
5 &
R \\ \addlinespace

Shoaib and Ramamohan (2022)\cite{shoaib_simulation_2022} &
\textit{Simulation} &
Resource utilisation in primary health centres in with differing resources, services and operational patterns &
India &
17 &
Python (\textit{salabim})\cite{van_der_ham_salabim_2018} \\ \addlinespace

Anagnostou et al. (2022)\cite{anagnostou_facs-charm_2022} &
\textit{Winter Simulation Conference} &
Intensive care unit capacity during COVID-19 &
Not specified &
1 &
Python (\textit{SimPy})\cite{team_simpy_simpy_2024} \\ \bottomrule

\end{tabular}}
\tabnote{* Scope refers to the number of items to be reproduced: figures, tables and/or results described in the text.}
\tabnote{** Language includes simulation package, where applicable; some studies instead developed the model from scratch.}
\tabnote{Abbreviations: COPD, chronic obstructive pulmonary disease; COVID-19, coronavirus disease 2019; UK, United Kingdom; US, United States.}
\label{table-studies}
\end{table}

\subsection{Reproduction}

Before starting, the authors of the eight studies were informed, and four were requested to add an open licence to their repository. A maximum of 40 hours was allowed for each reproduction (excluding computation time). The reproduction steps were as follows:

\begin{enumerate}
    \setlength\itemsep{1em}
    \item \textbf{Set-up.} A repository was created, including a logbook to track time and record progress and any barriers to running the code and reproducing results.
    \item \textbf{Scope.} At least two team members reviewed the article and agreed on which experimental results (e.g. tables, figures, results described in the text) to reproduce. The repository was then archived on Zenodo.\cite{european_organization_for_nuclear_research_zenodo_2013}
    \item \textbf{Running and troubleshooting.} A researcher set up an environment with the necessary software and packages. When creating a software environment for each reproduction, the protocol originally planned to backdate software and package versions to those specified or available at the time of publication. This approach was successfully implemented for Python studies. However, for studies using R, there were substantial challenges in backdating R and its packages. Hence, the latest versions were instead used, with code modifications applied as needed. Code was run and troubleshooted - typically by modifying or writing code, and consulting the article. Unresolved issues were discussed with the team, and then with the original authors, for whom response was optional. This approach differs from typical journal reproducibility assessments, as much more extensive troubleshooting was allowed in this research.
    \item \textbf{Reproduction success.} A consensus decision was made as to whether each item had been successfully reproduced or not. This was a subjective decision that allowed some expected deviation due to model stochasticity.
\end{enumerate}

\subsection{Evaluation}

Articles were evaluated using two DES reporting guidelines: STRESS-DES from Monks et al. (2019),\cite{monks_strengthening_2019} and the ``generic reporting checklist" from Zhang et al. (2020).\cite{zhang_reporting_2020} Repositories were evaluated against the STARS framework\cite{monks_towards_2024} and criteria for journal badges from: ACM,\cite{association_for_computing_machinery_acm_artifact_2020} National Information Standards Organisation (NISO),\cite{niso_reproducibility_badging_and_definitions_working_group_reproducibility_2021} Centre for Open Science (COS),\cite{blohowiak_badges_2023} Institute of Electrical and Electronics Engineers (IEEE),\cite{institute_of_electrical_and_electronics_engineers_ieee_about_2024} and \textit{Psychological Science}.\cite{association_for_psychological_science_aps_psychological_2024, hardwicke_transparency_2024} The evaluation was performed by one researcher, with consensus from a second for uncertain or unmet criteria. For reference, full guidelines and badge criteria are provided in Appendix \ref{appendix-criteria}.

\subsection{Synthesis and consolidation}

Results from the reproduction, evaluation, and reflections on the facilitators and barriers encountered were shared via a Quarto website\cite{allaire_quarto_2024} hosted on GitHub Pages.\cite{github_github_2024} The repository was organised into a research compendium,\cite{gentleman_statistical_2007} with a README, conventional file organisation (e.g. data, methods, outputs), tests, and a Docker environment.\cite{merkel_docker_2014} These tests, specifically written for this study, run the model to confirm it produces consistent results, either reproducing full paper outputs or, for long-running models, executing smaller verification cases. A second researcher verified execution and reproducibility by running the model and/or tests (locally and via Docker). The final repository was archived on Zenodo,\cite{european_organization_for_nuclear_research_zenodo_2013} and results were shared with the original authors.

\subsection{Citations and visualisations}

Studies are not referenced by name in the results, to keep the emphasis on overarching trends rather than individual reproduction success. For transparency, full results are available, with links to the reproduction and original study repositories provided in Appendix \ref{appendix-repo}. For this article, plots were created using Python 3.10.14,\cite{python_core_team_python_2024} \textit{Plotly} 5.23.0,\cite{plotly_technologies_inc_collaborative_2015} \textit{Matplotlib} 3.9.2\cite{hunter_matplotlib_2007} and \textit{pandas} 2.2.2,\cite{the_pandas_development_team_pandas-devpandas_2024,mckinney_data_2010} with a full list of dependencies in the research repository.\cite{heather_computational_2025} Other visualisations were created using Inkscape 1.3.2 (\url{https://inkscape.org/}) and Sketchpad v2022.2.21.0 (\url{https://sketch.io/sketchpad/}).

\section{Results}

\subsection{Reproduction}

\begin{figure}[!ht]
    \centering
    \resizebox*{14cm}{!}{\includegraphics{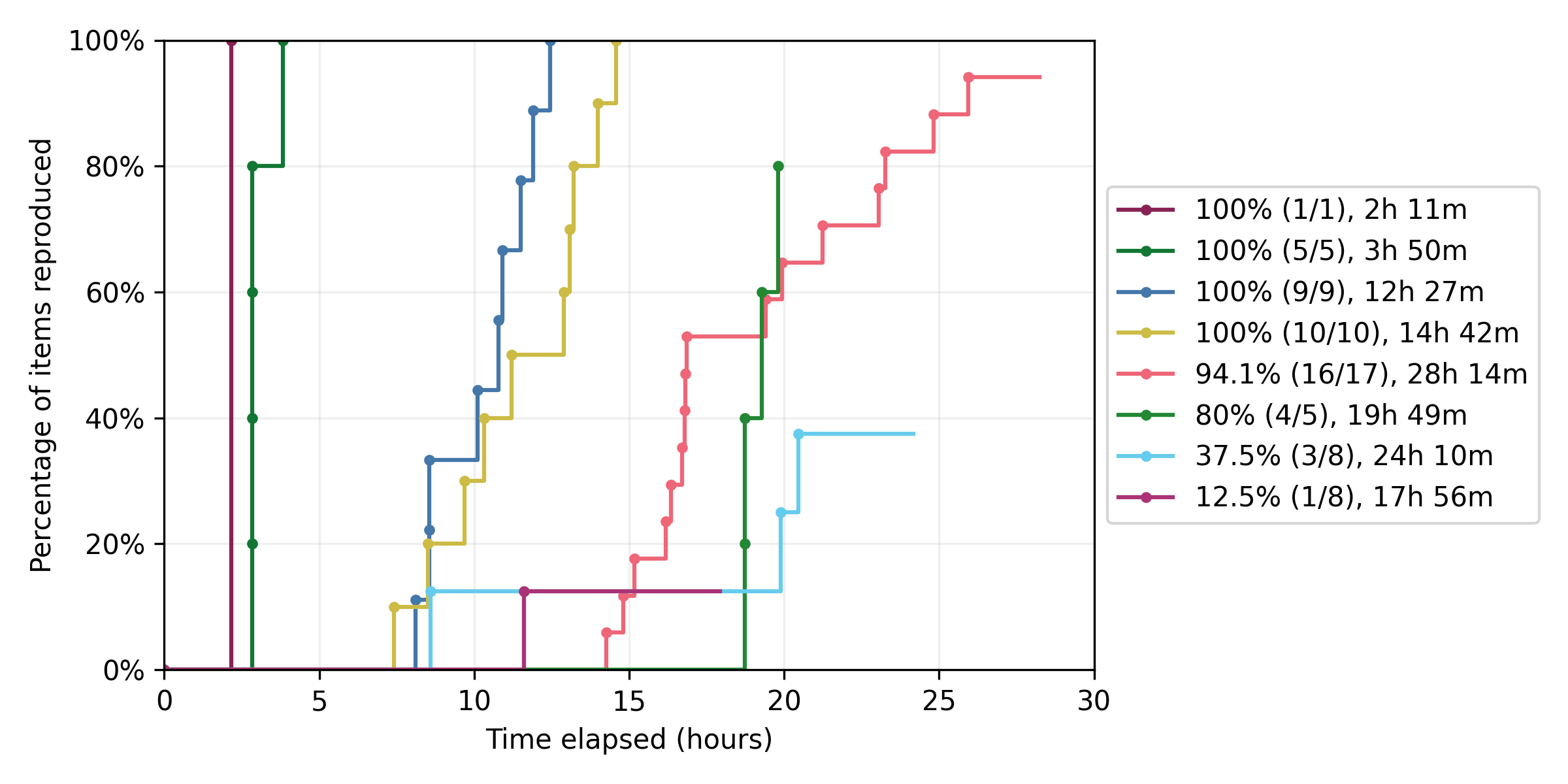}}
    \caption{Count, proportion, and time to reproduce items within the scope of each study. Inspired by a figure in Krafczyk et al. (2021)\cite{krafczyk_learning_2021}}
    \label{figure-times}
\end{figure}

Half of the studies were fully reproduced, with two completed in under four hours and two requiring 12 to 15 hours. The remaining four were partially reproduced (12.5\% to 94.1\%) and took 18 to 28 hours (Figure \ref{figure-times}). These times exclude model computation, which ranged widely: one model ran in seconds, three in under an hour, and four required several hours to days. Times were influenced by optimisations like parallel processing, and the machine used. Models were run on an Intel Core i7-12700H with 32GB RAM (Ubuntu 22.04.4), or an Intel Core i9-13900K with 81GB RAM (Pop!\_OS 22.04), which was able to run for longer amounts of time, and accommodated models with high memory demands.

\subsection{Evaluation of the repository}

\begin{figure}[!ht]
    \centering
    \resizebox*{14cm}{!}{\includegraphics{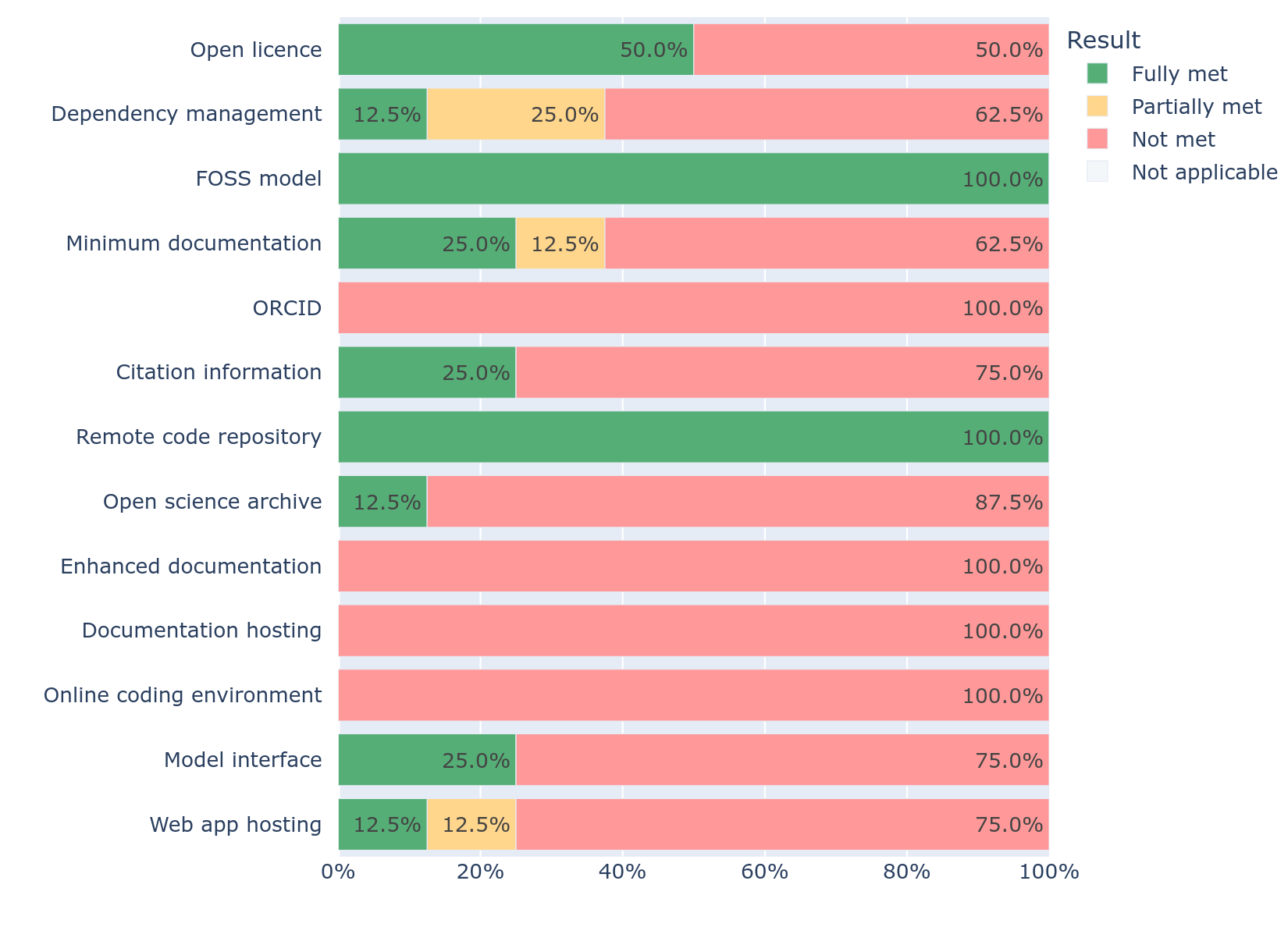}}
    \caption{Of the eight healthcare DES studies evaluated, the proportion that met each recommendation in the current STARS framework. For a full description of each criterion, see Appendix \ref{appendix-criteria}. \\ Abbreviations: DES, discrete-event simulation; STARS, \textit{Sharing Tools and Artefacts for Reusable Simulations}.}
    \label{figure-stars}
\end{figure}

Repositories were assessed using the STARS framework, with the results of the assessment in Figure \ref{figure-stars}. All models were FOSS and hosted on a remote repository, though only one was archived. Licensing was inconsistent, with half requiring follow-up with authors. Documentation, dependency management, citation information and ORCID were minimal. With regards to optional STARS components, two studies had applications, though none had enhanced documentation or online coding environments.

The repositories were also evaluated against journal badge criteria, with results for the ACM badges summarised here (Table \ref{table-acm}) and the full evaluation in Appendix \ref{appendix-badges}. Only one repository met the standards for the “Artefacts Available” badge. None satisfied “Artefacts Functional” requirements, typically due to incomplete code (missing scenarios, outputs, etc.). Whilst those evaluations used the original artefacts as shared by the study authors, the ``Results Reproduced" assessment was based on reproduction within a reasonable time allowing minimal troubleshooting, as per ACM criteria. This differs from the extended reproduction attempts described above, which allowed many hours of extensive troubleshooting. Under the stricter ACM criteria, only one study was eligible for ``Results Reproduced".

The proportion of STARS criteria met by each individual study ranged from 25\% to 88\% for essential components and 0\% to 40\% for optional components. These metrics showed no relationship with reproduction success (Table \ref{table-criteria}). However, some individual STARS components and badge criteria were critical to reproduction. For example, identifying the packages and versions necessary to run the code without error can be time-consuming if not stated by the authors, and dependency management is a component of both STARS and several badges. A complete set of materials, required for the ACM ``Artefacts Evaluated" badge (though absent in STARS), was also essential. Missing materials - such as parameters, scenarios, or code - were a major barrier to reproduction in most studies. Documentation is required by STARS and several badges, and were helpful if provided as they can reduce time spent deciphering how to run the code. Finally, an open licence, as specified in both STARS and for some badges, is fundamental to enable legal access and use of the code.

\begin{table}[!ht]
\tbl{Evaluation of repositories against ACM badge criteria.}
{\begin{tabular}{R{4cm} R{6cm} R{3cm}} \toprule
\textbf{Badge} & \textbf{Criteria} & \textbf{Studies that met criteria} \\ \midrule

ACM ``Artefacts Available" & \begin{minipage}[t]{\linewidth}\begin{itemize}[nosep, wide=0pt, leftmargin=*, after=\strut] 
    \item Artefacts are archived in a repository that is: (a) public (b) guarantees persistence (c) gives a unique identifier (e.g. DOI)
\end{itemize} \end{minipage} & 1/8 (12.5\%) \\ \addlinespace

ACM ``Artefacts Evaluated - Functional" & \begin{minipage}[t]{\linewidth} \begin{itemize}[nosep, wide=0pt, leftmargin=*, after=\strut] 
\item Documents (a) inventory of artefacts (b) sufficient description for artefacts to be exercised
\item Artefacts relevant to the paper
\item Complete (all relevant artefacts available)
\item Scripts can be successfully executed
\end{itemize} \end{minipage} & 0/8 (0.0\%) \\ \addlinespace

ACM ``Artefacts Evaluated - Reusable" & \begin{minipage}[t]{\linewidth}
Criteria for ``Functional" badge plus: 
\begin{itemize}[nosep, leftmargin=1.5em, after=\strut]
    \item Artefacts are carefully documented and well-structured to the extent that reuse and repurposing is facilitated, adhering to norms and standards
\end{itemize} \end{minipage} & 0/8 (0.0\%) \\ \addlinespace

ACM ``Results Reproduced" & \begin{minipage}[t]{\linewidth} \begin{itemize}[nosep, wide=0pt, leftmargin=*, after=\strut] 
\item Reproduced results (assuming (a) acceptably similar (b) reasonable time frame (c) only minor troubleshooting)
\end{itemize} \end{minipage} & 1/8 (12.5\%) \\ \bottomrule

\end{tabular}}
\tabnote{Abbreviations: ACM, Association for Computing Machinery; DOI, digital object identifier.}
\label{table-acm}
\end{table}

\vspace{1cm}

\begin{table}[!ht]
\tbl{The proportion of applicable criteria that were fully met, from evaluation of repository or article,  alongside the proportion of items reproduced from each study.}
{\begin{tabular}{R{3.4cm} R{2.4cm} R{2.4cm} R{2.4cm} R{2.4cm}} \toprule
\textbf{Items reproduced} & \textbf{STARS (essential)} & \textbf{STARS (optional)} & \textbf{STRESS-DES} & \textbf{Generic checklist} \\ \midrule
\multicolumn{5}{l}{\textbf{Fully reproduced}} \tabularnewline[5pt]
\hspace{3mm} 100\% (10/10) & 50\% & 0\% & 68\% & 71\% \tabularnewline[5pt]
\hspace{3mm} 100\% (9/9)   & 25\% & 0\% & 71\% & 75\% \tabularnewline[5pt]
\hspace{3mm} 100\% (5/5)   & 25\% & 0\% & 92\% & 76\% \tabularnewline[5pt]
\hspace{3mm} 100\% (1/1)   & 88\% & 20\% & 67\% & 53\% \tabularnewline[5pt] 
\multicolumn{5}{l}{\textbf{Partially reproduced}} \tabularnewline[5pt]
\hspace{3mm} 94.1\% (16/17)  & 25\% & 0\% & 71\% & 73\% \tabularnewline[5pt]
\hspace{3mm} 80\% (4/5)    & 50\% & 0\% & 84\% & 88\% \tabularnewline[5pt]
\hspace{3mm} 37.5\% (3/8)  & 25\% & 40\% & 61\% & 44\% \tabularnewline[5pt]
\hspace{3mm} 12.5\% (1/8)  & 25\% & 0\% & 78\% & 59\% \\ \bottomrule
\end{tabular}}
\tabnote{Abbreviations: STARS, \textit{Sharing Tools and Artefacts for Reusable Simulations}; STRESS-DES, \textit{Strengthening The Reporting of Empirical Simulation Studies - Discrete-Event Simulation}.}
\label{table-criteria}
\end{table}

\subsection{Evaluation of the article}

\begin{figure}[!htp]
    \centering
    \resizebox*{14cm}{!}{\includegraphics{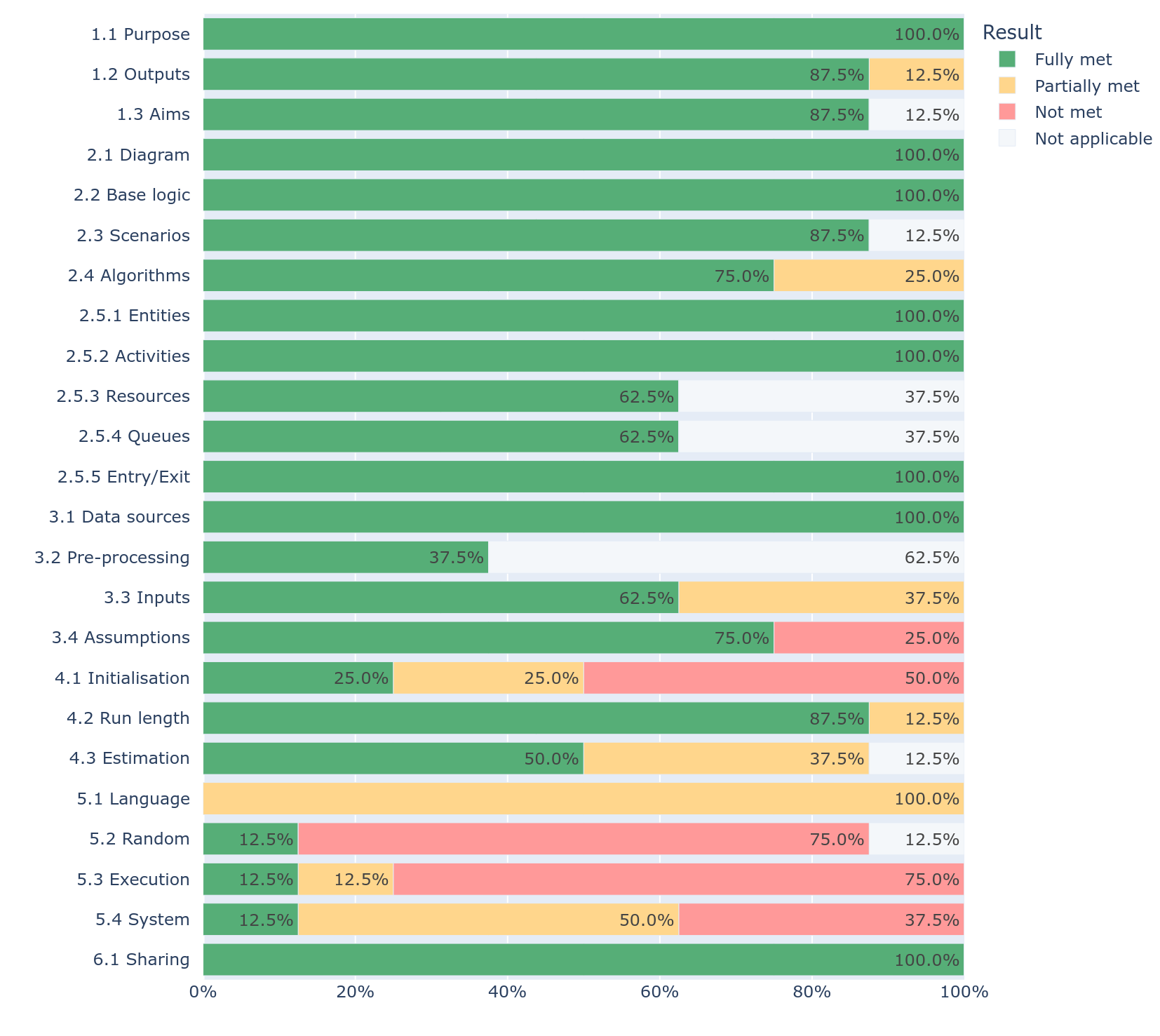}}
    \caption{Of the eight healthcare DES studies evaluated, the proportion that met each item in the current STRESS-DES criteria.\cite{monks_strengthening_2019} For a full description of each criterion, see Appendix \ref{appendix-criteria}. \\ Abbreviations: DES, discrete-event simulation; STRESS-DES, \textit{Strengthening The Reporting of Empirical Simulation Studies - Discrete-Event Simulation}.}
    \label{figure-stress}
\end{figure}

\begin{figure}[!htp]
    \centering
    \resizebox*{14cm}{!}{\includegraphics{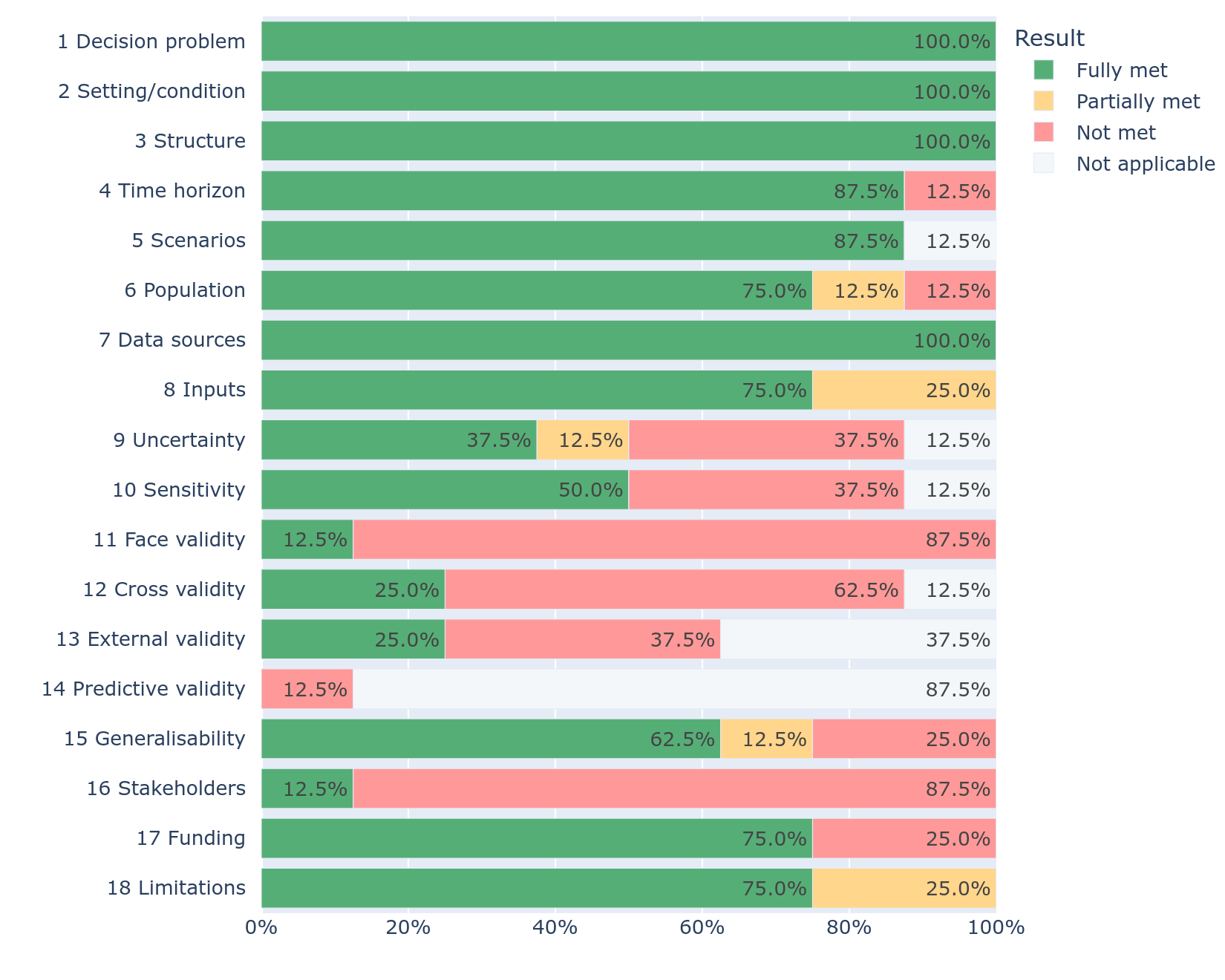}}
    \caption{Of the eight healthcare DES studies evaluated, the proportion that met each criteria in the general reporting checklist for DES.\cite{zhang_reporting_2020} For full description of each criteria, see Appendix \ref{appendix-criteria}. \\ Abbreviations: DES, discrete-event simulation.}
    \label{figure-ispor}
\end{figure}

Studies were evaluated against STRESS-DES\cite{monks_strengthening_2019} and the generic reporting checklist\cite{zhang_reporting_2020} - with the proportion of studies meeting each individual criterion presented in Figures \ref{figure-stress} and \ref{figure-ispor}, respectively. Most met criteria related to purpose and design. However, technical details in STRESS-DES - such as input parameters, initialisation, random sampling and execution - were often incomplete or unclear. While all studies mentioned the software or programming language used, these descriptions were minimal - for example, often mentioning the programming language and main simulation package, but not the operating system or versions used. System specifications were rarely fully described, and the requirement for pre-processing was unclear, and often marked as non-applicable. For the generic checklist, few studies addressed model uncertainties or performed sensitivity analyses, and validation was rarely discussed. Generalisability and stakeholder involvement were also seldom reported.

The proportion of applicable criteria met by each individual study is presented in Table \ref{table-criteria}. This ranged from 61\% to 92\% of the criteria per study for STRESS-DES and from 44\% to 88\% for the generic checklist. Although these values again showed no relationship with reproduction success (Table \ref{table-criteria}), certain elements were important to reproducibility. Complete and clear provision of input parameters (STRESS-DES 3.3, generic checklist 8) enabled verification and correction of parameters in the code. When code for the scenarios was missing, a clear description in the article was essential (STRESS-DES 2.3, generic checklist 5) - ideally specifying input parameters for all scenarios, as made explicit in STRESS-DES. Clear documentation of software and package versions (STRESS-DES 5.1) was helpful when dependency management was lacking. Run-time details (STRESS-DES 5.4) were also valuable, especially when studies had long execution times. When code for output calculations was missing, clear descriptions in the article were essential for understanding.

\section{Recommendations}

\begin{figure}[!ht]
    \centering
    \resizebox*{10cm}{!}{\includegraphics{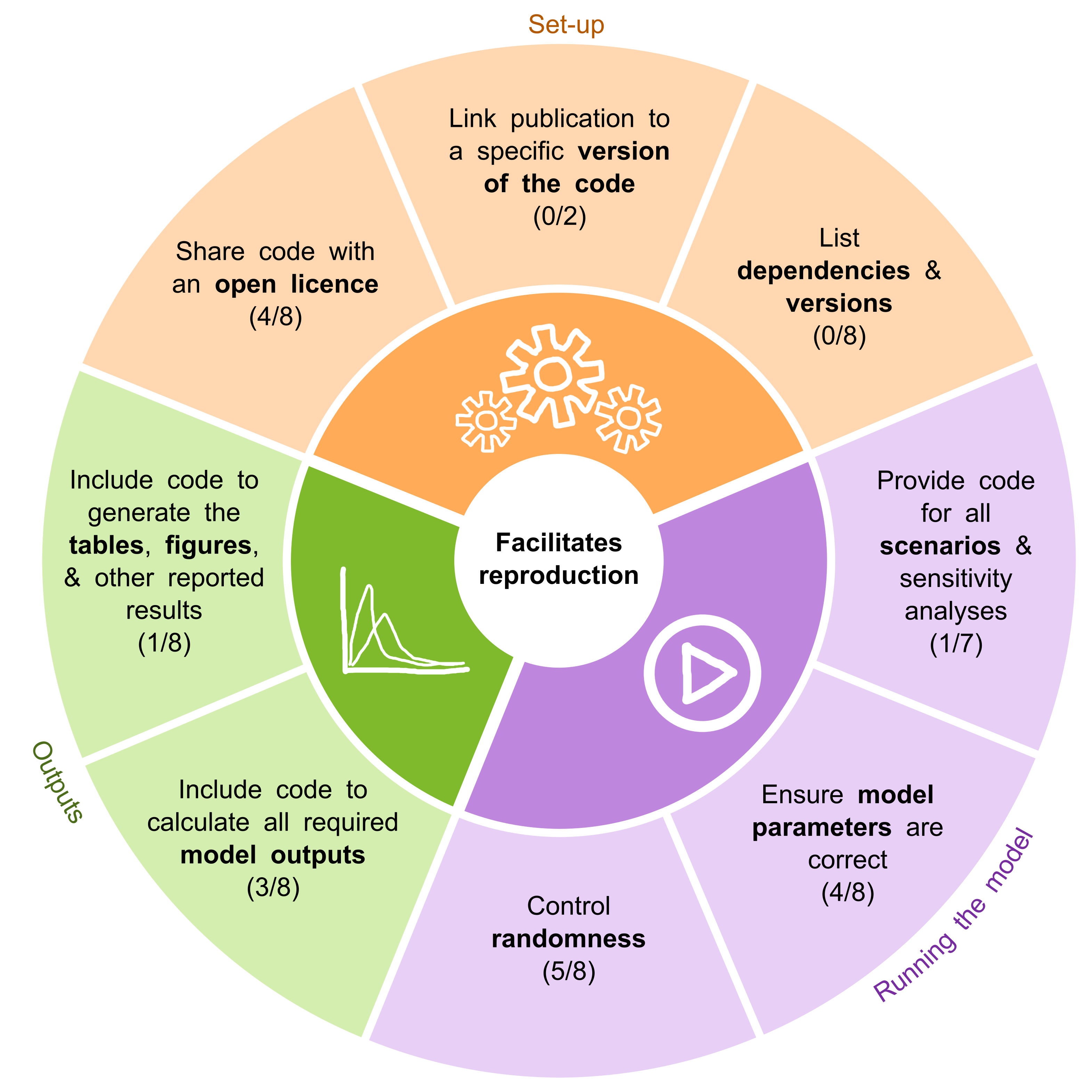}}
    \caption{Recommendations to support reproducibility. Below each recommendation, a count of studies that fully met it is provided. The total may fall below eight if the criteria were not applicable to a given study (e.g. if they didn't perform scenario analysis, or only provided one version of the code).}
    \label{figure-reproduction}
\end{figure}

\begin{figure}[!ht]
    \centering
    \resizebox*{10cm}{!}{\includegraphics{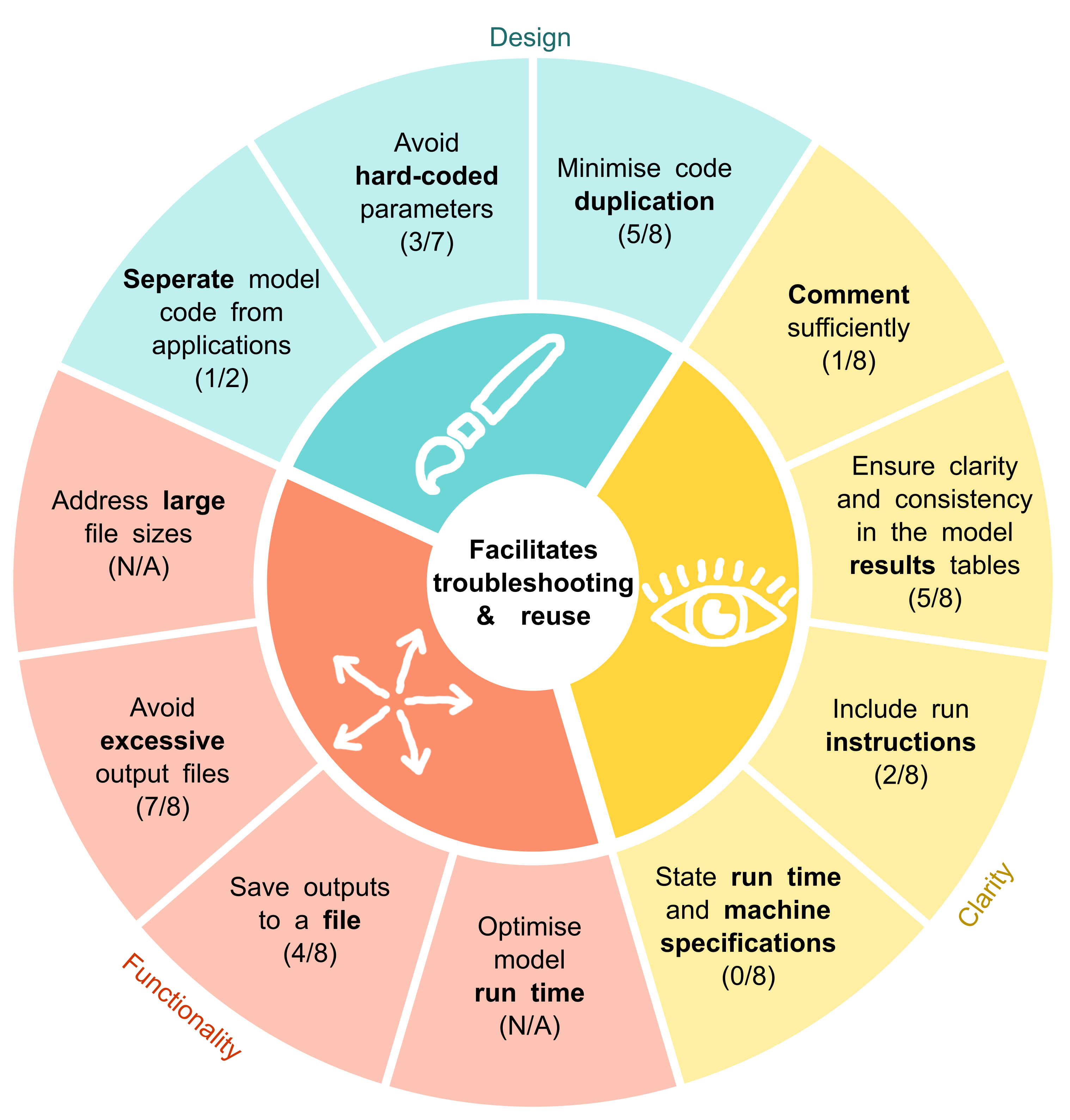}}
    \caption{Recommendations to support troubleshooting and reuse. Below each recommendation, a count of studies that fully met it is provided. The total may fall below eight if the criteria were not applicable to a given study (e.g. if they didn't have a web application, or didn't have scenarios to vary parameters). Some recommendations were marked as ``N/A" where it was not felt appropriate or feasible to count/assess their inclusion.\\ Abbreviations: N/A, not applicable.}
    \label{figure-reuse}
\end{figure}

The facilitators and barriers encountered during the reproducibility assessments are presented as a series of recommendations, grouped into two themes: factors that primarily supported reproduction itself, ensuring the provided code could be run and yielded the same results as reported in the articles (Figure \ref{figure-reproduction}), and factors that facilitated troubleshooting, enhancing the code’s adaptability for new contexts (Figure \ref{figure-reuse}).

\subsection{Recommendations to support reproducibility}

\subsubsection{Set-up}

\textbf{Share code with an open licence.} Half of the models were initially unlicensed, effectively preventing the code from being downloaded or run. Guidance on selecting an appropriate licence can be found in Morin et al. (2012)\cite{morin_quick_2012} and on \url{https://choosealicense.com/}.

\textbf{Link publication to a specific version of the code.} In one study, updates to the code after publication meant it was unclear which version produced the reported results. Journal articles can cite the exact version of code used to generate results if it is deposited in a open science archive. These archives, such as Figshare\cite{digital_science_figshare_2025} or Zenodo,\cite{european_organization_for_nuclear_research_zenodo_2013} provide a digital object identifier (DOI) for citation.\cite{van_gulick_resources_2024} Modified versions of code can be deposited at key stages during the publication process (e.g., submission, after revision, and acceptance) to obtain a new DOI. Code repositories can also contain a changelog (e.g. \url{https://keepachangelog.com}) that can help track changes between versions.

\textbf{List all dependencies and versions.} Only two studies provided a complete list of required packages, and none gave all versions. Environment managers like \textit{Conda}\cite{conda_contributors_conda_2023} or \textit{renv}\cite{ushey_renv_2024} are recommended for creating isolated environments and tracking dependencies. This includes recording the specific versions used, as code may break with package updates.

\subsubsection{Running the model}

\textbf{Provide code for all scenarios and sensitivity analyses.} Six of the seven studies with scenarios omitted code for these, often just providing the base case. Without code, it was challenging and time-consuming to implement the scenarios - especially as descriptions of the scenarios were often ambiguous, with the required parameters sometimes unclear or not stated.

\textbf{Ensure model parameters are correct.} In half the studies, code parameters did not align with those in the articles, often causing large differences in the model outcomes. It is important to ensure that the correct parameters are provided in the code and clearly documented in the article (which can be used to verify or correct the code).

\textbf{Control randomness.} Random seeds were not included in three studies. This made it impossible to determine whether differences between the results obtained from running the code and the results reported in the articles were due to variability in the simulation or issues in the code. Including a random seed is crucial for ensuring consistent results from stochastic models like DES. Tools like \texttt{set.seed()} in R\cite{r_core_team_r_2024} or \textit{NumPy's}\cite{harris_array_2020} \texttt{SeedSequence()} and \texttt{default\_rng()} are recommended.

\subsubsection{Outputs}

\textbf{Include code to calculate all required model outputs.} In five studies, some outputs were not generated by the provided code or included in the model's output tables. These included basic measures (eg. outcome counts), results from additional time points, and complex transformations of the provided outputs.

\textbf{Include code to generate the tables, figures, and other reported results}. Only one study provided code to generate all results in the article. Without this code, recreating outputs is time-consuming and challenging, requiring the identification of relevant results tables and columns, appropriate pre-processing steps, and the creation of the desired plots or tables.

\subsection{Recommendations to support troubleshooting and reuse}

\subsubsection{Design}

\textbf{Separate model code from applications.} One study embedded the model within the code for a web application (which itself did not produce the paper outputs). To execute it, the model had to be extracted from the app - so sharing standalone model scripts would be preferable.

\textbf{Avoid hard-coded parameters.} In four studies, parameters that varied between scenarios were hard-coded into model functions, making them difficult to adjust. Parameters should be defined as adjustable inputs, allowing the same code to run both base case and scenario analyses without modifications. This avoids duplicating files and ensures clear documentation of changes between scenarios.

\textbf{Minimise code duplication.} Three studies contained repeated sections of code, such as for each scenario, or model warm-up and execution. This duplication reduced code readability and increased the risk of errors - for example, when modifying parameters that were defined in multiple locations, and so had to be updated in each instance. It can be mitigated by reusing code through functions or classes.

\subsubsection{Clarity}

\textbf{Comment sufficiently.} Most studies were felt to have insufficient comments, hindering understanding of the parameters and code functionality when troubleshooting. Researchers should follow best practices, such as using docstrings for all functions and classes and adding concise yet informative comments elsewhere. Style guides, like PEP-8 (\url{https://peps.python.org/pep-0008/}) and PEP-257 (\url{https://peps.python.org/pep-0257/}) in Python, and the tidyverse style guide (\url{https://style.tidyverse.org/}) in R, offer recommendations on writing code documentation.

\textbf{Ensure clarity and consistency in the model results tables.} In some studies, there was uncertainty regarding which tables, columns, or scenarios to use from the model results. For instance, one model produced two alternative spreadsheets with overlapping metrics but occasionally differing results, making it unclear which set of results was correct. In another case, it was not initially evident that a reported result was derived from a combination of columns, rather than from a single column. To address these issues, columns should be clearly labelled, results should be consistent across tables, and data dictionaries should be provided for complex outputs. Additionally, it should be explicitly clear which elements contribute to the results, ideally by including the relevant code.

\textbf{Include run instructions.} Only two studies provided guidance. Clear instructions, including script run order and any necessary commands, are essential - especially in complex analyses with several scripts.

\textbf{State run times and machine specifications.} Clearly stating the expected run times and specifications of the machines used helps users understand the resource demands and assess if their machine can handle the workload, as they may not have the capacity for long or memory-intensive runs. Times can be easily recorded such as by using the \textit{time} module from Python's standard library,\cite{python_core_team_python_2024} or the \texttt{Sys.time()} function from base R.\cite{r_core_team_r_2024} It can also be helpful to record the memory usage when running the model.

\subsubsection{Functionality}

\textbf{Optimise model run time.} Long run times were a significant challenge in this research. Reducing run times is valuable both during initial testing, where quicker runs aid troubleshooting and debugging, and for full-scale runs, which may be impractical if requiring machines be left running continuously for extended periods. Options such as simplified configurations designed for model verification (e.g., reducing entities or replications) and parallel processing could significantly reduce run times or enable minimal test runs for troubleshooting and/or development.

\textbf{Save outputs to a file}. In four studies, results were output to dataframes within the Python/R environment but not saved to files. Without saving, the entire model must be re-executed to modify analyses, which is inefficient, especially for long-running models. Saving outputs allows for easier manipulation and analysis in separate scripts.

\textbf{Avoid excessive output files.} In one study, the model generated numerous files which were not used in analysis. Researchers should implement a run mode that limits output to essential files for analysis, while still allowing additional files for verification or debugging. Use a \texttt{.gitignore} file to prevent uploading unnecessary files to GitHub.

\textbf{Address large file sizes.} Output files sometimes exceeded GitHub's 100MB threshold. This can be addressed by compressing files (e.g. \texttt{.csv.gz}) or using GitHub Large File Storage - although that has limits under GitHub's free tier.

\subsection{Key recommendations}

From the complete set of recommendations, five were identified as having the greatest impact on the reproductions, based on both fundamental enablers of reproduction and observed patterns of what consistently created barriers.

\begin{enumerate}
    \item \textbf{Share code with an open licence.}
    \item \textbf{Ensure model parameters are correct.}
    \item \textbf{Include code to calculate all required model outputs.}
    \item \textbf{Provide code for all scenarios and sensitivity analyses.}
    \item \textbf{Include code to generate the tables, figures, and other reported results.}
\end{enumerate}

The first recommendation, sharing code with an open licence, is fundamental because, without it, others cannot legally use, modify or share the code, making reproduction otherwise impossible. The remaining recommendations (items 2 to 5) have been prioritised because they consistently posed significant barriers to achieving similar results to the original article, and/or required extensive time to troubleshoot and resolve.

\section{Discussion}

This research identified factors affecting reproducibility in healthcare DES models and proposed five key recommendations to address these issues. These recommendations were highlighted based on their critical impact, either because they were essential prerequisites for reproduction, caused significant discrepancies in reproduced results, or required extensive troubleshooting. In this discussion, these recommendations are reflected upon in the context of evidence from other studies and established best practices.

First, sharing models with an open licence. Half of the models in this research (50\%, n=4/8) initially lacked a licence, similar to findings in broader healthcare DES reviews (65\%, n=26/47)\cite{monks_computer_2023} and other fields, such as computational physics (71\%, n=39/55) and \textit{Science} articles (34\%, n=19/56).\cite{stodden_empirical_2018} Open licensing is recommended by initiatives like the NHS ``Goldacre review",\cite{goldacre_better_2022} journal badge programmes,\cite{association_for_computing_machinery_acm_artifact_2020, niso_reproducibility_badging_and_definitions_working_group_reproducibility_2021,blohowiak_badges_2023} and the STARS framework.\cite{monks_towards_2024} While private code can support internal reproducibility, it limits broader validation and reuse.

Second, maintaining an accurate and consistent record of the model parameters between the article and code was critical, as incorrect parameters often had a large impact on the observed outcomes. Third, sharing code for scenarios and sensitivity analyses - not just the base case - was vital, as reconstructing these without access to the original code was time-consuming and error-prone. These recommendations align with requirements from frameworks like STRESS-DES\cite{monks_strengthening_2019} and the generic reporting checklist,\cite{zhang_reporting_2020} as well as prior research identifying missing data, parameters, and scripts as major barriers to reproducibility across disciplines.\cite{stodden_empirical_2018, konkol_computational_2019, fisar_reproducibility_2024, mccullough_lessons_2006, stodden_enabling_2018, krafczyk_learning_2021, obels_analysis_2020}

The fourth and fifth recommendations involve sharing code to calculate all required model outputs and to then generate the results presented in the article (e.g. tables, figures). A lack of visualisation code or reliance on proprietary visualisation tools has similarly hindered reproduction in computational physics.\cite{krafczyk_learning_2021, stodden_enabling_2018} These recommendations, along with others, can help produce a reproducible analytical pipeline (RAP), wherein executing the code will run the model and produce all results from the paper. Tools like Python and R facilitate RAPs by generating figures and tables directly from code, avoiding the manual input required in drag-and-drop tools. RAPs have been highlighted as a priority by initiatives like the NHS RAP Community of Practice\cite{nhs_england_rap_community_of_practice_reproducible_2024} and the UK Government Analysis Function RAP Strategy.\cite{reproducible_analytical_pipelines_rap_team_reproducible_2022} More widely, initiatives such as the European Commission's exploration of reproducibility challenges in scientific research\cite{baker_reproducibility_2020} and a European Union project focused on RAPs for monitoring plastic pollution\cite{community_research_and_development_information_service_cordis_european_2024} reflect the growing international emphasis on reproducibility and RAPs in diverse contexts.

In addition to the five key recommendations, further recommendations - such as setting seeds, linking to specific code versions, listing dependencies, avoiding hard coding, adding comments, providing run instructions, minimising unnecessary outputs, clarifying results tables, and considering runtime and hardware needs - are also backed by the literature as important reproducibility factors across various fields.\cite{eubank_lessons_2016, krafczyk_learning_2021, konkol_computational_2019, mccullough_lessons_2006, stodden_empirical_2018, obels_analysis_2020}

Some reproducibility issues are difficult to avoid, even with thorough preparation, as they may depend on future software updates or system differences. For example, one of the models used an older, now unsupported Python version (2.7.12), which may not work with some code editors/interpreters. Creating environments with an older language was straightforward in Python via \textit{Conda}\cite{conda_contributors_conda_2023} but was more challenging and not successful in R in this research. R packages are intended to be forward compatible, although this cannot be guaranteed, with newer packages breaking the code in some studies.\cite{trisovic_large-scale_2022, konkol_computational_2019} Similarly, one study shared a web application but this was no longer accessible remotely after its hosting site shut down. System-level code dependencies can also vary depending on the operating system used, and though they can be identified by testing the code on fresh builds, this may be beyond the scope of many projects.

\subsection{Implications}

\subsubsection{For model developers}

Model developers, including researchers and practitioners, should consider adopting the outlined recommendations to improve reproducibility, which can, in turn, help facilitate model reuse. The relevance of this is highlighted by the fact that three of the eight articles analysed in this research reused models described in prior studies:

\begin{itemize}
    \item Kim et al. (2021)\cite{kim_modelling_2021} used a model described in Glover et al. (2018)\cite{glover_discrete_2018} and Thompson et al. (2018).\cite{thompson_screening_2018}
    \item Johnson et al. (2021)\cite{johnson_cost_2021} used a model described in Sadatsafavi et al. (2019).\cite{sadatsafavi_development_2019}
    \item Wood et al. (2021)\cite{wood_value_2021} used a model described in Wood et al. (2020).\cite{wood_covid-19_2020}
\end{itemize}

Common concerns about the skills\cite{eynden_survey_2016, hrynaszkiewicz_survey_2021} or time\cite{eynden_survey_2016, cadwallader_survey_2022, borghi_data_2021, hrynaszkiewicz_survey_2021} required to prepare before sharing code can be mitigated by making a few incremental improvements throughout the project. Four of the key recommendations in this research - using correct parameters, running scenarios, producing outputs, and creating tables and figures - will already have been performed by researchers throughout the research process, and so the primary action is simply to share these complete artefacts. The fifth recommendation, adding an open licence, is a simple but essential step, enabling others to download, run, and use the code. Other recommendations from this article may require a greater time investment, though this can be minimised if considered from the outset of the project. The more recommendations that are implemented, the greater the anticipated improvement in the reproducibility of the research.

\subsubsection{For peer reviewers}

Regardless of whether journal or conference policies are in place, reviewers can promote reproducibility by incorporating a ``reproducibility review" (Table \ref{table-review}). This proposed review incorporates three of the five main recommendations from this research: ensuring open licences, availability of scenario and sensitivity analysis code, and providing result-processing scripts. These checks are straightforward to identify from any provided code or artefacts. Alternatively, they could be posed as peer review questions that an author could self-certify. The remaining two recommendations—verifying consistent parameters and confirming that the model generates all outputs—were excluded, as they require a much greater time investment to investigate, including potentially needing to run the code. This review is consistent with the requirements for openly licensed and complete materials from ACM’s ``Artefacts Evaluated" \cite{association_for_computing_machinery_acm_artifact_2020} and IEEE’s ``Code Reviewed" \cite{institute_of_electrical_and_electronics_engineers_ieee_about_2024} badges. The review complements other existing suggestions, such as ``open scholarship review" and ``longevity review" (Table \ref{table-review}) from Monks and Harper (2023). \cite{monks_computer_2023}

\begin{table}[!ht]
\tbl{Simple checklists to assist reviewers in assessing the openness, longevity, and reproducibility of DES models during peer review.}
{\begin{tabular}{R{2.25cm} R{8.5cm} R{2.25cm}} \toprule
\textbf{Review} & \textbf{Checklist} & \textbf{Proposer}\\ \midrule

Open scholarship review &
\begin{minipage}[t]{\linewidth}
\begin{itemize}[nosep, wide=0pt, leftmargin=*, after=\strut] 
    \item Used a reporting checklist?
    \item Deposited model in a public archive?
    \item Included ORCID in model metadata?
    \item Shared model under an open licence?
    \item Provided basic instructions to run and use model?
\end{itemize} \end{minipage}
& Monks and Harper (2023)\cite{monks_computer_2023} \\ \addlinespace

Longevity review &
\begin{minipage}[t]{\linewidth}
\begin{itemize}[nosep, wide=0pt, leftmargin=*, after=\strut] 
    \item Used a reporting checklist?
    \item Explained how dependencies are managed?
\end{itemize} \end{minipage}
& Monks and Harper (2023)\cite{monks_computer_2023} \\ \addlinespace

Reproducibility review &
\begin{minipage}[t]{\linewidth}
\begin{itemize}[nosep, wide=0pt, leftmargin=*, after=\strut] 
    \item Shared model under an open licence?
    \item Provided code for all scenarios and sensitivity analyses?
    \item Provided code for tables, figures and other reported results?
\end{itemize} \end{minipage}
& Present study \\ \bottomrule

\end{tabular}}
\tabnote{Abbreviations: ORCID, Open Researcher and Contributor Identifier.}
\label{table-review}
\end{table}

\subsubsection{For reporting guidelines}

Levels of adherence to the reporting guidelines were similar to those observed in prior studies. The mean proportion of fully met items on the generic reporting checklist\cite{zhang_reporting_2020} in this research - either calculated out of applicable criteria (67.4\%) or all criteria (61.1\%) - was similar to previous evaluations of 211 healthcare DES articles (63.7\%)\cite{zhang_reporting_2020} and 18 radiotherapy pathway simulation articles (63.6\%).\cite{robinson_simulation_2023} Gaps in uncertainty assessment, sensitivity analysis, generalisability and validity were common across the studies.\cite{zhang_reporting_2020, robinson_simulation_2023} Although they don't report on individual items, a previous study did assess adherence to STRESS-DES for 13 pharmacoeconomic models, finding higher mean adherence (80.4\%)\cite{nwanosike_direct_2023} than in this study - as calculated out of applicable criteria (74.0\%) or all criteria (68.2\%).

In this research, adherence to the reporting checklists did not translate to reproducibility; as observed in prior studies with other checklists.\cite{schwander_replication_2021,mcmanus_barriers_2019,stodden_enabling_2018} This is unsurprising as the generic checklist emphasises model quality whilst STRESS-DES is focused on reporting all necessary information to facilitate replication of simulations. However, some items on these checklists - like describing input parameters, scenarios and dependencies - were helpful in the reproductions. Focusing on STRESS-DES, this could be adapted to further facilitate reproducibility:

\begin{enumerate}
    \item \textbf{Clarification:} Provide suggestions to improve clarity in the reporting of model outputs, scenarios and input parameters (e.g. using tables) - since these were often partially or ambiguously described.
    \item \textbf{Simplification:} Break down complex sections into smaller, more manageable items for easier reporting and review. For example, section 5.3 ``Model execution" \cite{monks_strengthening_2019} covers several distinct topics in only one item: the event processing mechanism; priority rules for entities competing for resources; use of parallel, distributed and cloud computing; time management algorithms; and details about high-level architecture (version, run-time, supporting documents).
    \item \textbf{Redundancy:} Reconsider sections on random number generation and event-processing mechanisms, as modern tools may not need this level of detail.
    \item \textbf{Sharing:} Recommend including the completed checklist as supplementary material, which would provide direct access to relevant information (which can be tricky to find in complex articles - up to two hours per evaluation in this research).
\end{enumerate}

Completion and sharing of reporting checklists as supplementary material - though not guaranteed to facilitate reproducibility - certainly increase the transparency of reported models. Alongside the checklists in this research, others exist for different model types - for example, STRESS-ABS for agent-based simulation and STRESS-SD for system dynamics models,\cite{monks_strengthening_2019} as well as the ODD protocol (\textit{Overview, Design concepts and Details protocol for describing Individual and Agent-Based Models}).\cite{grimm_odd_2020} These checklists are complemented by TRACE (\textit{TRAnsparent and Comprehensive Ecological modelling documentation}), a framework for documenting model development, testing and analysis\cite{grimm_towards_2014} - supported by maintaing modelling notebooks through the development process.\cite{ayllon_keeping_2021}

\subsubsection{For the STARS framework}

The STARS framework was designed to support healthcare DES model reuse. As it was published in 2024,\cite{monks_towards_2024} it was not used in the reviewed studies. This research examined whether incidental adherence to its recommendations improved reproducibility; no clear link was found. This is similar to findings in other fields where practices like documentation and archiving do not guarantee reproducibility.\cite{stodden_enabling_2018, henderson_reproducibility_2024} This suggests that, while these practices are valuable, other factors contribute to reproducibility challenges. Modifications to STARS that may better facilitate reproducibility include requiring a complete set of materials (i.e. parameters, scenarios, results processing), which aligns with key recommendations from this study and existing badge expectations. Additionally, incorporating guidelines on runtime/hardware, as well as code design (e.g. seeds, running scenarios programmatically, docstrings/comments, output saving).

\subsection{Strengths and limitations}

This study is the first to systematically assess the computational reproducibility of healthcare DES models, providing insights that are broadly relevant to other contexts and modelling approaches. A key strength is the transparency of the research, with all code openly licensed, available on GitHub, archived on Zenodo (Appendix \ref{appendix-repo}), and supported by a pre-registered protocol.\cite{heather_protocol_2024} Additionally, a complementary website summarises the results and provides further details (\url{https://pythonhealthdatascience.github.io/stars_wp1_summary/}).\cite{heather_computational_2025} The study benefits from a diverse selection of eight studies, although they are biased toward models with shared licensing and resources, which are uncommon in healthcare DES.\cite{monks_computer_2023} The reproducibility process was time-intensive, requiring up to 28 hours per study, which may not reflect typical researcher resources - but this time investment was crucial to understanding the factors influencing reproducibility, requiring a close examination of the troubleshooting for each case.

Reproductions were conducted by a single researcher, occasionally consulting others when troubleshooting. If the aim of this research had been to determine the exact level of reproducibility, it would have been necessary for multiple researchers to independently attempt to reproduce the studies, to ensure reliable results. A prior study of psychology articles found low inter-rater reliability in such assessments (75\% agreement on the executability of R scripts and 56\% agreement on reproducibility).\cite{obels_analysis_2020} However, it was felt suitable for a single researcher to conduct the reproductions, given the study’s focus on identifying factors impacting reproducibility, rather than definitive reproduction rates. Barriers and facilitators encountered are expected to be relatively consistent between researchers - as evidenced by the complementary recommendations from reproductions in other fields. Determining successful reproduction involved subjective judgement - acknowledged in the protocol\cite{heather_protocol_2024} - so decisions were made by consensus between at least two team members. 

The study focused on models implemented in Python and R, excluding those developed using commercial DES software. Although most DES studies use commercial off-the-shelf software, the majority (62\%) of open DES models use FOSS \cite{monks_computer_2023}. Our expectation is that reproduction of results generated by commercial software may present some differences - particularly in troubleshooting the unique nuances of commercial DES software as well as access to a compatible and suitably licensed version. However, we also expect the top five study recommendations to remain broadly applicable.  For example, a simulation model will still need a licence (CC-BY may be best suited to a simulation file); and correct parameters will still be required.  

While absolute and percentage differences between the reproductions and original results were considered, these could vary significantly based on scale. For example, percentage differences are greater for smaller values (e.g., 0.1, 0.2) than for larger values (e.g., 10, 15), even if the latter are more practically significant. The nature of the metrics analysed could also impact outcomes. For example, the costs and Quality-Adjusted Life Years from a health economic model in this study appeared similar, but even small differences led to substantial variations in derived calculations (e.g. Incremental Cost-Effectiveness Ratio).

Evaluations were conducted by one researcher, with a second checking uncertain or unmet criteria, and all evaluations were revisited side-by-side for consistency. The potential for mistakes in the evaluations highlights the value of attaching completed reporting guidelines to articles to make technical details more accessible — particularly for longer papers or when information is spread across multiple appendices or prior publications.

\section{Conclusion}

This study identifies critical factors impacting the reproducibility of healthcare DES models. It offers actionable recommendations to improve the reproducibility of shared models, emphasising five key recommendations: adopting an open licence, ensuring model parameters are correct, ensuring the model outputs all required metrics, including code to run all scenarios and/or sensitivity analyses, and including code to generate all tables, figures and other reported results. While adherence to guidelines like STRESS-DES and STARS did not translate to successful reproductions, suggested improvements to these frameworks could help facilitate reproducibility. Further work could include refining these frameworks and providing practical examples of RAP in DES that meet these criteria, as examples could help address concerns about the skills and time investment needed for implementation.

\section{Acknowledgements}

We would like to thank the authors who made their code available under open licences, facilitating our research. We are especially grateful to the following authors for their helpful communication during the project: Ivan Hernandez, Tze Ping Loh, Chun Yee Lim, Lois Kim, Kate Johnson, Richard Wood, Mohd Shoaib, and Anastasia Anagnostou. The circular design of Figures 7 and 8 was inspired by Gomes et al. (2022).\cite{gomes_why_2022}

\section{Data availability statement}

The records supporting the findings of this study - including code, models, results and other relevant materials - are synthesised and available at \url{https://pythonhealthdatascience.github.io/stars_wp1_summary/} and archived on Zenodo.\cite{heather_computational_2025} Links to the individual repositories for the reproduced studies and the original studies are provided in Appendix \ref{appendix-repo}.

\section{Funding}

This work was supported by the Medical Research Council under grant number MR/Z503915/1.

\section{Disclosure of interest}

The authors report there are no competing interests to declare.

\pagebreak
\appendix

\section{Frameworks and guidelines used in evaluation}\label{appendix-criteria}

\subsection{STRESS-DES}

As presented in Monks et al. (2019).\cite{monks_strengthening_2019}

\begin{itemize}
    \item \textbf{1.1 Purpose} - Purpose of the model - Explain the background and objectives for the model.
    \item \textbf{1.2 Outputs} - Model outputs - Define all quantitative performance measures that are reported, using equations where necessary. Specify how and when they are calculated during the model run along with how any measures of error such as confidence intervals are calculated.
    \item \textbf{1.3 Aims} - Experimentation aims - If the model has been used for experimentation, state the objectives that it was used to investigate. 
    \begin{itemize}
        \item (A) Scenario-based analysis – Provide a name and description for each scenario, providing a rationale for the choice of scenarios and ensure that item 2.3 (below) is completed.
        \item (B) Design of experiments – Provide details of the overall design of the experiments with reference to performance measures and their parameters (provide further details in data below).
        \item (C) Simulation Optimisation – (if appropriate) Provide full details of what is to be optimised, the parameters that were included and the algorithm(s) that was be used. Where possible provide a citation of the algorithm(s).
    \end{itemize}
    \item \textbf{2.1 Diagram} - Base model overview diagram - Describe the base model using appropriate diagrams and description. This could include one or more process flow, activity cycle or equivalent diagrams sufficient to describe the model to readers. Avoid complicated diagrams in the main text. The goal is to describe the breadth and depth of the model with respect to the system being studied.
    \item \textbf{2.2 Base logic} - Base model logic - Give details of the base model logic. Give additional model logic details sufficient to communicate to the reader how the model works.
    \item \textbf{2.3 Scenarios} - Scenario logic - Give details of the logical difference between the base case model and scenarios (if any). This could be incorporated as text or where differences are substantial could be incorporated in the same manner as 2.2.
    \item \textbf{2.4 Algorithms} - Algorithms - Provide further detail on any algorithms in the model that (for example) mimic complex or manual processes in the real world (i.e. scheduling of arrivals/ appointments/ operations/ maintenance, operation of a conveyor system, machine breakdowns, etc.). Sufficient detail should be included (or referred to in other published work) for the algorithms to be reproducible. Pseudo-code may be used to describe an algorithm.
    \item \textbf{2.5.1 Entities} - Components - entities - Give details of all entities within the simulation including a description of their role in the model and a description of all their attributes.
    \item \textbf{2.5.2 Activities} - Components - activities - Describe the activities that entities engage in within the model. Provide details of entity routing into and out of the activity.
    \item \textbf{2.5.3 Resources} - Components - resources - List all the resources included within the model and which activities make use of them.
    \item \textbf{2.5.4 Queues} - Components - queues - Give details of the assumed queuing discipline used in the model (e.g. First in First Out, Last in First Out, prioritisation, etc.). Where one or more queues have a different discipline from the rest, provide a list of queues, indicating the queuing discipline used for each. If reneging, balking or jockeying occur, etc., provide details of the rules. Detail any delays or capacity constraints on the queues.
    \item \textbf{2.5.5 Entry/exit} - Components - entry/exit points - Give details of the model boundaries i.e. all arrival and exit points of entities. Detail the arrival mechanism (e.g. ‘thinning’ to mimic a non-homogenous Poisson process or balking).
    \item \textbf{3.1 Data sources} - Data sources - List and detail all data sources. Sources may include: interviews with stakeholders, samples of routinely collected data, prospectively collected samples for the purpose of the simulation study, public domain data published in either academic or organisational literature. Provide, where possible, the link and digital object identifier (DOI) to the data or reference to published literature. All data source descriptions should include details of the sample size, sample date ranges and use within the study.
    \item \textbf{3.2 Pre-processing} - Pre-processing - Provide details of any data manipulation that has taken place before its use in the simulation, e.g. interpolation to account for missing data or the removal of outliers.
    \item \textbf{3.3 Inputs} - Input parameters - List all input variables in the model. Provide a description of their use and include parameter values. For stochastic inputs provide details of any continuous, discrete or empirical distributions used along with all associated parameters. Give details of all time-dependent parameters and correlation. Clearly state: 
    \begin{itemize}
        \item Base case data.
        \item Data use in experimentation, where different from the base case.
        \item Where optimisation or design of experiments has been used, state the range of values that parameters can take.
        \item Where theoretical distributions are used, state how these were selected and prioritised above other candidate distributions.
    \end{itemize}
    \item \textbf{3.4 Assumptions} - Assumptions - Where data or knowledge of the real system is unavailable what assumptions are included in the model? This might include parameter values, distributions or routing logic within the model.
    \item \textbf{4.1 Initialisation} - Initialisation - Report if the system modelled is terminating or non-terminating. State if a warm-up period has been used, its length and the analysis method used to select it. For terminating systems state the stopping condition. State what if any initial model conditions have been included, e.g., pre-loaded queues and activities. Report whether initialisation of these variables is deterministic or stochastic.
    \item \textbf{4.2 Run length} - Run length - Detail the run length of the simulation model and time units.
    \item \textbf{4.3 Estimation} - Estimation approach - State the method used to account for the stochasticity: For example, two common methods are multiple replications or batch means. Where multiple replications have been used, state the number of replications and for batch means, indicate the batch length and whether the batch means procedure is standard, spaced or overlapping. For both procedures provide a justification for the methods used and the number of replications/size of batches.
    \item \textbf{5.1 Language} - Software or programming language - State the operating system and version and build number. State the name, version and build number of commercial or open source discrete-event simulation (DES) software that the model is implemented in. State the name and version of general-purpose programming languages used (e.g. Python 3.5). Where frameworks and libraries have been used provide all details including version numbers.
    \item \textbf{5.2 Random} - Random sampling - State the algorithm used to generate random samples in the software/programming language used e.g. Mersenne Twister. If common random numbers are used, state how seeds (or random number streams) are distributed among sampling processes.
    \item \textbf{5.3 Execution} - Model execution - State the event processing mechanism used e.g. three phase, event, activity, process interaction. Note that in some commercial software the event processing mechanism may not be published. In these cases authors should adhere to item 5.1 software recommendations. State all priority rules included if entities/activities compete for resources. If the model is parallel, distributed and/or use grid or cloud computing, etc., state and preferably reference the technology used. For parallel and distributed simulations the time management algorithms used. If the High Level Architecture (HLA) is used then state the version of the standard, which run-time infrastructure (and version), and any supporting documents (Federation Object Models (FOMs), etc.).
    \item \textbf{5.4 System} - System specification State the model run time and specification of hardware used. This is particularly important for large scale models that require substantial computing power. For parallel, distributed and/or use grid or cloud computing, etc. state the details of all systems used in the implementation (processors, network, etc.).
    \item \textbf{6.1 Sharing} - Computer model sharing statement - Describe how someone could obtain the model described in the paper, the simulation software and any other associated software (or hardware) needed to reproduce the results. Provide, where possible, the link and DOIs to these.
\end{itemize}

\subsection{Generic reporting checklist}

As outlined in Zhang et al. (2020).\cite{zhang_reporting_2020} \\

\noindent \textbf{Model conceptualisation:}

\begin{itemize}
    \item \textbf{1 Decision problem} - Is the focused health-related decision problem clarified? ...the decision problem under investigation was defined. DES studies included different types of decision problems, eg, those listed in previously developed taxonomies.
    \item \textbf{2 Setting/condition} - Is the modeled healthcare setting/health condition clarified? ...the physical context/scope (eg, a certain healthcare unit or a broader system) or disease spectrum simulated was described.
    \item \textbf{3 Structure} - Is the model structure described? ...the model's conceptual structure was described in the form of either graphical or text presentation.
    \item \textbf{4 Time horizon} - Is the time horizon given? ...the time period covered by the simulation was reported.
    \item \textbf{5 Scenarios} - Are all simulated strategies/scenarios specified? ...the comparators under test were described in terms of their components, corresponding variations, etc.
    \item \textbf{6 Population} - Is the target population described? ...the entities simulated and their main attributes were characterised.
\end{itemize}

\vspace{1cm}

\noindent \textbf{Parameterisation and uncertainty assessment:}

\begin{itemize}
    \item \textbf{7 Data sources} - Are data sources informing parameter estimations provided? ...the sources of all data used to inform model inputs were reported.
    \item \textbf{8 Inputs} - Are the parameters used to populate model frameworks specified? ...all relevant parameters fed into model frameworks were disclosed.
    \item \textbf{9 Uncertainty} - Are model uncertainties discussed? ...the uncertainty surrounding parameter estimations and adopted statistical methods (eg, 95\% confidence intervals or possibility distributions) were reported.
    \item \textbf{10 Sensitivity} - Are sensitivity analyses performed and reported? ...the robustness of model outputs to input uncertainties was examined, for example via deterministic (based on parameters’ plausible ranges) or probabilistic (based on a priori-defined probability distributions) sensitivity analyses, or both.
\end{itemize}

\noindent \textbf{Validation:}

\begin{itemize}
    \item \textbf{11 Face validity} - Is face validity evaluated and reported? ...it was reported that the model was subjected to the examination on how well model designs correspond to the reality and intuitions. It was assumed that this type of validation should be conducted by external evaluators with no stake in the study.
    \item \textbf{12 Cross validity} - Is cross validation performed and reported? ...comparison across similar modelling studies which deal with the same decision problem was undertaken.
    \item \textbf{13 External validity} - Is external validation performed and reported? ...the modeler(s) examined how well the model’s results match the empirical data of an actual event modeled.
    \item \textbf{14 Predictive validity} - Is predictive validation performed or attempted? ...the modeler(s) examined the consistency of a model’s predictions of a future event and the actual outcomes in the future. If this was not undertaken, it was assessed whether the reasons were discussed.
\end{itemize}

\noindent \textbf{Generalisability and stakeholder involvement:}

\begin{itemize}
    \item \textbf{15 Generalisability} - Is the model generalizability issue discussed? ...the modeler(s) discussed the potential of the resulting model for being applicable to other settings/populations (single/multiple application).
    \item \textbf{16 Stakeholders} - Are decision makers or other stakeholders involved in modelling? ...the modeler(s) reported in which part throughout the modelling process decision makers and other stakeholders (eg, subject experts) were engaged.
    \item \textbf{17 Funding} - Is the source of funding stated? ...the sponsorship of the study was indicated.
    \item \textbf{18 Limitations} - Are model limitations discussed? ...limitations of the assessed model, especially limitations of interest to decision makers, were discussed.
\end{itemize}

\subsection{STARS framework}

As described in Monks et al. (2024).\cite{monks_towards_2024} \\

\noindent \textbf{Essential components:}

\begin{itemize}
    \item \textbf{Open licence} - Free and open-source software (FOSS) licence (e.g. MIT, GNU Public Licence (GPL)).
    \item \textbf{Dependency management} - Specify software libraries, version numbers and sources (e.g. dependency management tools like virtualenv, conda, poetry).
    \item \textbf{FOSS model} - Coded in FOSS language (e.g. R, Julia, Python).
    \item \textbf{Minimum documentation} - Minimal instructions (e.g. in README) that overview (a) what model does, (b) how to install and run model to obtain results, and (c) how to vary parameters to run new experiments.
    \item \textbf{ORCID} - ORCID for each study author.
    \item \textbf{Citation information} - Instructions on how to cite the research artefact (e.g. CITATION.cff file).
    \item \textbf{Remote code repository} - Code available in a remote code repository (e.g. GitHub, GitLab, BitBucket).
    \item \textbf{Open science archive} - Code stored in an open science archive with FORCE11 compliant citation and guaranteed persistance of digital artefacts (e.g. Figshare, Zenodo, the Open Science Framework (OSF), and the Computational Modelling in the Social and Ecological Sciences Network (CoMSES Net)).
\end{itemize}

\noindent \textbf{Optional components:}

\begin{itemize}
    \item \textbf{Enhanced documentation} - Open and high quality documentation on how the model is implemented and works (e.g. via notebooks and markdown files, brought together using software like Quarto and Jupyter Book). Suggested content includes: 
    \begin{itemize}
        \item Plain english summary of project and model.
        \item Clarifying licence.
        \item Citation instructions.
        \item Contribution instructions.
        \item Model installation instructions.
        \item Structured code walk through of model.
        \item Documentation of modelling cycle using \textit{TRAnsparent and Comprehensive model Evaluation} (TRACE).
        \item Annotated simulation reporting guidelines.
        \item Clear description of model validation including its intended purpose.
    \end{itemize}
    \item \textbf{Documentation hosting} - Host documentation (e.g. with GitHub pages, GitLab pages, BitBucket Cloud, Quarto Pub).
    \item \textbf{Online coding environment} - Provide an online environment where users can run and change code (e.g. BinderHub, Google Colaboratory, Deepnote).
    \item \textbf{Model interface} - Provide web application interface to the model so it is accessible to less technical simulation users.
    \item \textbf{Web app hosting} - Host web app online (e.g. Streamlit Community Cloud, ShinyApps hosting).
\end{itemize}

\subsection{Journal badges}

Journal badges were identified from several sources:

\begin{itemize}
    \item Association for Computing Machinery (ACM).\cite{niso_reproducibility_badging_and_definitions_working_group_reproducibility_2021}
    \item National Information Standards Organisation (NISO).\cite{niso_reproducibility_badging_and_definitions_working_group_reproducibility_2021}
    \item Center for Open Science (COS).\cite{blohowiak_badges_2023}
    \item Institute of Electrical and Electronics Engineers (IEEE).\cite{institute_of_electrical_and_electronics_engineers_ieee_about_2024}
    \item \textit{Psychological Science} journal.\cite{association_for_psychological_science_aps_psychological_2024, hardwicke_transparency_2024}
\end{itemize}

The criteria for each badge are described in Table \ref{table-badge_criteria}, with the badges grouped into three themes, as defined by NISO.\cite{niso_reproducibility_badging_and_definitions_working_group_reproducibility_2021}

\begin{table}[!htp]
\tbl{Criteria for journal badges.}
{\begin{tabular}{R{3.7cm} R{9.3cm}} \toprule
\textbf{Badge} & \textbf{Criteria} \\ \addlinespace \midrule

\textbf{``Open objects" badges} \\ \addlinespace

ACM ``Artefacts Available" \cite{association_for_computing_machinery_acm_artifact_2020} & \begin{minipage}[t]{\linewidth}\begin{itemize}[nosep, wide=0pt, leftmargin=*, after=\strut] 
    \item Artefacts are archived in a repository that is: (a) public (b) guarantees persistence (c) gives a unique identifier.
\end{itemize} \end{minipage} \\ \addlinespace

NISO ``Open Research Objects (ORO)" \cite{niso_reproducibility_badging_and_definitions_working_group_reproducibility_2021} & \begin{minipage}[t]{\linewidth}
Criteria for ACM badge plus:
\begin{itemize}[nosep, wide=0pt, leftmargin=*, after=\strut]
    \item Open licence.
\end{itemize} \end{minipage} \\ \addlinespace

NISO ``Open Research Objects - All (ORO-A)" \cite{niso_reproducibility_badging_and_definitions_working_group_reproducibility_2021} & \begin{minipage}[t]{\linewidth} 
Criteria for NISO ``ORO" badge plus: 
\begin{itemize}[nosep, leftmargin=1.5em, after=\strut]
    \item Complete (all relevant artefacts available).
\end{itemize} \end{minipage} \\ \addlinespace

COS ``Open Code" \cite{blohowiak_badges_2023} & \begin{minipage}[t]{\linewidth}
Criteria for NISO ``ORO" badge plus: 
\begin{itemize}[nosep, wide=0pt, leftmargin=*, after=\strut] 
    \item Documents (a) how code is used (b) how it relates to article (c) software, systems, packages and versions.
\end{itemize} \end{minipage} \\ \addlinespace

IEEE ``Code Available" \cite{institute_of_electrical_and_electronics_engineers_ieee_about_2024} & \begin{minipage}[t]{\linewidth} \begin{itemize}[nosep, wide=0pt, leftmargin=*, after=\strut] 
\item Complete (all relevant artefacts available).
\end{itemize} \end{minipage} \\ \addlinespace

\textbf{``Object review" badges} \\ \addlinespace

ACM ``Artefacts Evaluated - Functional" \cite{association_for_computing_machinery_acm_artifact_2020} & \begin{minipage}[t]{\linewidth} \begin{itemize}[nosep, wide=0pt, leftmargin=*, after=\strut] 
\item Documents (a) inventory of artefacts (b) sufficient description for artefacts to be exercised.
\item Artefacts relevant to the paper.
\item Complete (all relevant artefacts available).
\item Scripts can be successfully executed.
\end{itemize} \end{minipage} \\ \addlinespace

ACM ``Artefacts Evaluated - Reusable" \cite{association_for_computing_machinery_acm_artifact_2020} & \begin{minipage}[t]{\linewidth}
Criteria for ACM ``Functional" badge plus: 
\begin{itemize}[nosep, leftmargin=1.5em, after=\strut]
    \item Artefacts are carefully documented and well-structured to the extent that reuse and repurposing is facilitated, adhering to norms and standards.
\end{itemize} \end{minipage} \\ \addlinespace

IEEE ``Code Reviewed" \cite{institute_of_electrical_and_electronics_engineers_ieee_about_2024} & \begin{minipage}[t]{\linewidth} \begin{itemize}[nosep, wide=0pt, leftmargin=*, after=\strut] 
\item Complete (all relevant artefacts available).
\item Scripts can be successfully executed.
\end{itemize} \end{minipage} \\  \addlinespace

\textbf{``Reproduced" badges} \\ \addlinespace

ACM ``Results Reproduced" \cite{association_for_computing_machinery_acm_artifact_2020} & \begin{minipage}[t]{\linewidth} \begin{itemize}[nosep, wide=0pt, leftmargin=*, after=\strut] 
\item Reproduced results (assuming (a) acceptably similar (b) reasonable time frame (c) only minor troubleshooting).
\end{itemize} \end{minipage} \\ \addlinespace

NISO ``Results Reproduced (ROR-R)" \cite{niso_reproducibility_badging_and_definitions_working_group_reproducibility_2021} & Same as ACM ``Results Reproduced". \\ \addlinespace

IEEE ``Code Reproducible" \cite{winter_retrospective_2022} & Same as ACM ``Results Reproduced". \\ \addlinespace

Psychological Science ``Computational Reproducibility" \cite{association_for_psychological_science_aps_psychological_2024, hardwicke_transparency_2024} & \begin{minipage}[t]{\linewidth}
Criteria for ACM ``Results Reproduced" plus: 
\begin{itemize}[nosep, leftmargin=1.5em, after=\strut]
    \item README file with step-by-step instructions to run analysis.
    \item Dependencies (e.g. package versions) stated.
    \item Clear how output of analysis corresponds to article.
\end{itemize} \end{minipage} \\ \bottomrule
\end{tabular}}
\tabnote{Abbreviations: ACM, Association for Computing Machinery; COS, Center for Open Science; IEEE, Institute of Electrical and Electronics Engineers; NISO, National Information Standards Organisation.}
\label{table-badge_criteria}
\end{table}

\pagebreak
\section{Full journal badge evaluation}\label{appendix-badges}

\begin{table}[htp]
\tbl{Evaluation of studies against badge criteria.}
{\begin{tabular}{R{9cm} R{4cm}} \toprule
\textbf{Badge} & \textbf{Studies that met criteria} \\ \addlinespace \midrule

\textbf{``Open objects" badges} \\ \addlinespace

ACM ``Artefacts Available" \cite{association_for_computing_machinery_acm_artifact_2020} & 1/8 (12.5\%) \\ \addlinespace

NISO ``Open Research Objects (ORO)" \cite{niso_reproducibility_badging_and_definitions_working_group_reproducibility_2021} & 1/8 (12.5\%) \\ \addlinespace

NISO ``Open Research Objects - All (ORO-A)" \cite{niso_reproducibility_badging_and_definitions_working_group_reproducibility_2021} & 0/8 (0.0\%) \\ \addlinespace

COS ``Open Code" \cite{blohowiak_badges_2023} & 1/8 (12.5\%) \\ \addlinespace

IEEE ``Code Available" \cite{institute_of_electrical_and_electronics_engineers_ieee_about_2024} & 1/8 (12.5\%) \\ \addlinespace

\textbf{``Object review" badges} \\ \addlinespace

ACM ``Artefacts Evaluated - Functional" \cite{association_for_computing_machinery_acm_artifact_2020} & 0/8 (0.0\%) \\ \addlinespace

ACM ``Artefacts Evaluated - Reusable" \cite{association_for_computing_machinery_acm_artifact_2020} & 0/8 (0.0\%) \\ \addlinespace

IEEE ``Code Reviewed" \cite{institute_of_electrical_and_electronics_engineers_ieee_about_2024} & 1/8 (12.5\%) \\  \addlinespace

\textbf{``Reproduced" badges} \\ \addlinespace

ACM ``Results Reproduced" \cite{association_for_computing_machinery_acm_artifact_2020} & 1/8 (12.5\%) \\ \addlinespace

NISO ``Results Reproduced (ROR-R)" \cite{niso_reproducibility_badging_and_definitions_working_group_reproducibility_2021} & 1/8 (12.5\%) \\ \addlinespace

IEEE ``Code Reproducible" \cite{winter_retrospective_2022} & 1/8 (12.5\%) \\ \addlinespace

Psychological Science ``Computational Reproducibility" \cite{association_for_psychological_science_aps_psychological_2024, hardwicke_transparency_2024} & 0/8 (0.0\%) \\ \bottomrule
\end{tabular}}
\tabnote{Abbreviations: ACM, Association for Computing Machinery; COS, Center for Open Science; IEEE, Institute of Electrical and Electronics Engineers; NISO, National Information Standards Organisation.}
\label{table-badge_evaluation}
\end{table}

\pagebreak
\section{Code repositories}\label{appendix-repo}

The following table provides links to the code repositories containing each of the computational reproducibility assessments and evaluations. There are also links to the associated Quarto websites and archives, ensuring easy access to all relevant resources.


\begin{table}[!ht]
\tbl{Links for reproduction and evaluation.}{
\begin{tabular}{R{2cm} R{11cm}} \toprule
\textbf{Description} & \textbf{URL} \\ \midrule
\multicolumn{2}{l}{\textbf{Compilation of results from the eight studies}} \\
\quad Repository & \url{https://github.com/pythonhealthdatascience/stars_wp1_summary} \\
\quad Archive & \cite{heather_computational_2025} \\
\quad Website & \url{https://pythonhealthdatascience.github.io/stars_wp1_summary/} \\ \addlinespace
\multicolumn{2}{l}{\textbf{Reproduction: Hernandez et al. (2015) \cite{hernandez_optimal_2015}}} \\
\quad Repository & \url{https://github.com/pythonhealthdatascience/stars-reproduce-hernandez-2015} \\
\quad Archive & \cite{heather_hernandez_2025} \\
\quad Website & \url{https://pythonhealthdatascience.github.io/stars-reproduce-hernandez-2015/} \\ \addlinespace
\multicolumn{2}{l}{\textbf{Reproduction: Huang et al. (2019) \cite{huang_optimizing_2019}}} \\
\quad Repository & \url{https://github.com/pythonhealthdatascience/stars-reproduce-huang-2019} \\
\quad Archive & \cite{heather_huang_2025} \\
\quad Website & \url{https://pythonhealthdatascience.github.io/stars-reproduce-huang-2019/} \\ \addlinespace
\multicolumn{2}{l}{\textbf{Reproduction: Lim et al. (2020) \cite{lim_staff_2020}}} \\
\quad Repository & \url{https://github.com/pythonhealthdatascience/stars-reproduce-lim-2020} \\
\quad Archive & \cite{heather_lim_2025} \\
\quad Website & \url{https://pythonhealthdatascience.github.io/stars-reproduce-lim-2020/} \\ \addlinespace
\multicolumn{2}{l}{\textbf{Reproduction: Kim et al. (2021) \cite{kim_modelling_2021}}} \\
\quad Repository & \url{https://github.com/pythonhealthdatascience/stars-reproduce-kim-2021/} \\
\quad Archive & \cite{heather_kim_2025} \\
\quad Website & \url{https://pythonhealthdatascience.github.io/stars-reproduce-kim-2021/} \\ \addlinespace
\multicolumn{2}{l}{\textbf{Reproduction: Johnson et al. (2021) \cite{johnson_cost_2021}}} \\
\quad Repository & \url{https://github.com/pythonhealthdatascience/stars-reproduce-johnson-2021} \\
\quad Archive & \cite{heather_johnson_2025} \\
\quad Website & \url{https://pythonhealthdatascience.github.io/stars-reproduce-johnson-2021/} \\ \addlinespace
\multicolumn{2}{l}{\textbf{Reproduction: Wood et al. (2021) \cite{wood_value_2021}}} \\
\quad Repository & \url{https://github.com/pythonhealthdatascience/stars-reproduce-wood-2021} \\
\quad Archive & \cite{heather_wood_2025} \\
\quad Website & \url{https://pythonhealthdatascience.github.io/stars-reproduce-wood-2021/} \\ \addlinespace
\multicolumn{2}{l}{\textbf{Reproduction: Shoaib et al. (2022) \cite{shoaib_simulation_2022}}} \\
\quad Repository & \url{https://github.com/pythonhealthdatascience/stars-reproduce-shoaib-2022} \\
\quad Archive & \cite{heather_shoaib_2025} \\
\quad Website & \url{https://pythonhealthdatascience.github.io/stars-reproduce-shoaib-2022/} \\ \addlinespace
\multicolumn{2}{l}{\textbf{Reproduction: Anagnostou et al. (2022) \cite{anagnostou_facs-charm_2022}}} \\
\quad Repository & \url{https://github.com/pythonhealthdatascience/stars-reproduce-anagnostou-2022} \\
\quad Archive & \cite{heather_anagnostou_2025} \\
\quad Website & \url{https://pythonhealthdatascience.github.io/stars-reproduce-anagnostou-2022/} \\ \bottomrule
\end{tabular}}
\end{table}

\pagebreak

This second table provides links to the code repositories from the original articles.

\begin{table}[!ht]
\tbl{Links to original study repositories.}{
\begin{tabular}{R{2cm} R{11cm}} \toprule
\textbf{Description} & \textbf{URL} \\ \midrule
\multicolumn{2}{l}{\textbf{Original study: Hernandez et al. (2015) \cite{hernandez_optimal_2015}}} \\
\quad Repository & \url{https://github.com/ivihernandez/staff-allocation} \\ \addlinespace
\multicolumn{2}{l}{\textbf{Original study: Huang et al. (2019) \cite{huang_optimizing_2019}}} \\
\quad Repository & \url{https://github.com/shiweih/desECR} \\ \addlinespace
\multicolumn{2}{l}{\textbf{Original study: Lim et al. (2020) \cite{lim_staff_2020}}} \\
\quad Repository & \url{https://github.com/chaose5/COVID-roster-simulation} \\ \addlinespace
\multicolumn{2}{l}{\textbf{Original study: Kim et al. (2021) \cite{kim_modelling_2021}}} \\
\quad Repository & \url{https://github.com/mikesweeting/AAA_DES_model} \\ \addlinespace
\multicolumn{2}{l}{\textbf{Original study: Johnson et al. (2021) \cite{johnson_cost_2021}}} \\
\quad Repository & \url{https://github.com/KateJohnson/epicR/tree/closed_cohort} \\ \addlinespace
\multicolumn{2}{l}{\textbf{Original study: Wood et al. (2021) \cite{wood_value_2021}}} \\
\quad Repository & \url{https://github.com/nhs-bnssg-analytics/triage-modelling} \\ \addlinespace
\multicolumn{2}{l}{\textbf{Original study: Shoaib et al. (2022) \cite{shoaib_simulation_2022}}} \\
\quad Repository & \url{https://github.com/shoaibiocl/PHC-} \\ \addlinespace
\multicolumn{2}{l}{\textbf{Original study: Anagnostou et al. (2022) \cite{anagnostou_facs-charm_2022}}} \\
\quad Repository & \url{https://gitlab.com/anabrunel/charm} \\
\quad Archive & \cite{anagnostou_charm_2022} \\ \bottomrule
\end{tabular}}
\end{table}

\pagebreak

\bibliography{references.bib}

\end{document}